\documentclass{aa}

\usepackage{lscape}
\usepackage{amsfonts,amsmath,natbib,xcolor,subfigure,gensymb,longtable,ulem}
\usepackage{xtab,booktabs}
\usepackage{hyperref}
\hypersetup{breaklinks,colorlinks,citecolor=blue,linkcolor=blue,urlcolor=red}

\usepackage{graphicx}
\usepackage{txfonts}

\usepackage{orcidlink}  

\defcitealias{Reichart01}{R01}
\defcitealias{Guidorzi05b}{G05}

\begin{document}

\title{New results on the gamma-ray burst variability--luminosity relation\thanks{Tables~\ref{longtab:VL0.80_0.6} to \ref{longtab:VL0.45_0.6} are only available in electronic form at the CDS via anonymous ftp to cdsarc.u-strasbg.fr (130.79.128.5) or via http://cdsweb.u-strasbg.fr/cgi-bin/qcat?J/A+A/}}
\author{C.~Guidorzi\thanks{guidorzi@fe.infn.it}\inst{\ref{unife},\ref{infnfe},\ref{inafbo}}\orcidlink{0000-0001-6869-0835}
\and R.~Maccary\inst{\ref{unife},\ref{inafbo}}\orcidlink{0000-0002-8799-2510}
\and A.~Tsvetkova\inst{\ref{unica},\ref{ioffe}}\orcidlink{0000-0003-0292-6221}
\and S.~Kobayashi\inst{\ref{ljmu}}\orcidlink{0000-0001-7946-4200}
\and L.~Amati\inst{\ref{inafbo}}\orcidlink{0000-0001-5355-7388}
\and L.~Bazzanini\inst{\ref{unife},\ref{inafbo}}\orcidlink{0000-0003-0727-0137}
\and M.~Bulla\inst{\ref{unife},\ref{infnfe},\ref{inafte}}\orcidlink{0000-0002-8255-5127}
\and A.~E.~Camisasca\inst{\ref{oavda}}\orcidlink{0000-0002-4200-1947}
\and L.~Ferro\inst{\ref{unife},\ref{inafbo}}\orcidlink{0009-0006-1140-6913}
\and D.~Frederiks\inst{\ref{ioffe}}\orcidlink{0000-0002-1153-6340}
\and F.~Frontera\inst{\ref{unife},\ref{inafbo}}\orcidlink{0000-0003-2284-571X}
\and A.~Lysenko\inst{\ref{ioffe}}\orcidlink{0000-0002-3942-8341}
\and M.~Maistrello\inst{\ref{unife},\ref{inafbo}}\orcidlink{0009-0000-4422-4151}
\and A.~Ridnaia\inst{\ref{ioffe}}\orcidlink{0000-0001-9477-5437}
\and D.~Svinkin\inst{\ref{ioffe}}\orcidlink{0000-0002-2208-2196}
\and M.~Ulanov\inst{\ref{ioffe}}\orcidlink{0000-0002-0076-5228}
}

\institute{Department of Physics and Earth Science, University of Ferrara, Via Saragat 1, I-44122 Ferrara, Italy \label{unife}
\and  INFN -- Sezione di Ferrara, Via Saragat 1, 44122 Ferrara, Italy\label{infnfe}
\and INAF -- Osservatorio di Astrofisica e Scienza dello Spazio di Bologna, Via Piero Gobetti 101, 40129 Bologna, Italy \label{inafbo}
\and Department of Physics, University of Cagliari, SP Monserrato-Sestu, km $0.7$, 09042 Monserrato, Italy \label{unica}
\and Ioffe Institute, Politekhnicheskaya 26, 194021 St. Petersburg, Russia \label{ioffe}
\and Astrophysics Research Institute, Liverpool John Moores University, Liverpool Science Park IC2, 146 Brownlow Hill, Liverpool,  L3 5RF, UK \label{ljmu}
\and INAF, Osservatorio Astronomico d'Abruzzo, via Mentore Maggini snc, 64100 Teramo, Italy \label{inafte}
\and Astronomical Observatory of the Autonomous Region of the Aosta Valley (OAVdA), Loc. Lignan 39, I-11020, Nus (Aosta Valley), Italy\label{oavda}
}
\abstract
{At the dawn of the gamma--ray burst (GRB) afterglow era, a Cepheid-like correlation was discovered between time variability $V$ and isotropic-equivalent peak luminosity $L_{\rm iso}$ of the prompt emission of about a dozen long GRBs with measured redshift available at that time. Soon afterwards, the correlation was confirmed against a sample of about 30 GRBs, despite being affected by significant scatter. Unlike the minimum variability timescale (MVT), $V$ measures the relative power of short-to-intermediate timescales.}
{We aim to test the correlation using about two hundred long GRBs with spectroscopically measured redshift, detected by {\it Swift}, {\it Fermi}, and Konus/{\it WIND}, for which both observables can be accurately estimated.}
{For all the selected GRBs, variability was calculated according to the original definition using the 64-ms background-subtracted light curves of {\it Swift}/BAT ({\it Fermi}/GBM) in the 15--150 (8--900) keV energy passband. Peak luminosities were either taken from literature or derived from modelling broad-band spectra acquired with either Konus/{\it WIND} or {\it Fermi}/GBM.}
{The statistical significance of the correlation has weakened to $\lesssim 2$\%, mostly due to the appearance of a number of smooth and luminous GRBs characterised by a relatively small $V$. At odds with most long GRBs, 3 out of 4 long-duration merger candidates have high $V$ and low $L_{\rm iso}$.}
{Luminosity is more tightly connected with shortest timescales measured by MVT rather than short-to-intermediate ones, measured by $V$. We discuss the implications on internal dissipation models and the role of the $e^{\pm}$ photosphere. We identified a few, smooth GRBs with a single broad pulse and low $V$, that might have an external shock origin, in contrast with most GRBs. The combination of high variability $(V\gtrsim 0.1)$, low luminosity $L_{\rm iso}\lesssim 10^{51}$~erg~s$^{-1}$ and short MVT ($\lesssim 0.1$~s) could be a good indicator for a compact binary merger origin.}

\maketitle

\section{Introduction}
\label{sec:intro}
In the early days of multi-messenger and time-domain astrophysics and big surveys, the number and diversity of explosive transients are rapidly increasing and are expected to continue doing so in the coming years.
In this context, gamma-ray bursts (GRBs) remain central to the field, due to the expanding sample of electromagnetic detections at TeV energies (see \citealt{Nava21,NodaParsons22} for reviews), their association with gravitational waves (GWs; \citealt{LIGO-Fermi17}), and their potential as sources of high-energy neutrinos (see \citealt{Meszaros17_GRBneutrinos,MuraseBartos19,Kimura23} for reviews). 

Since their first discovery, the wealth of information encoded in GRB prompt emission light curves (LCs) has provided valuable insights, ranging from the classification of progenitors to constraints on the dissipation mechanism and radii. Regarding the progenitors, short GRBs signal the merger of a compact object binary system \citep{Blinnikov1984, Eichler89,Paczynski91,Narayan92,LIGO-Fermi17}, while long GRBs indicate the core collapse of certain massive stars also known as ``collapsar''  \citep{Woosley93,Paczynski98,MacFadyen99,Yoon05}. Although recent cases have shown that duration alone can be misleading, other indicators related to the temporal properties of the prompt emission may aid in their identification \citep{Camisasca23,Veres23}.

At the dawn of the afterglow era, the diversity and complexity of long GRB LCs led to the development of various metrics to quantify their variability. The common goal was to evaluate how a given LC fluctuates around a smoothed version of itself, emphasising the relative power of short-to-intermediate timescales compared to long ones (\citealt{Fenimore00}; \citealt[hereafter R01]{Reichart01}). The rationale was that the power associated with the shortest timescales (down to a few ten ms and only rarely to ms, \citealt{Golkhou15}), which modulates the profiles of long and very long GRBs without apparent evolution, supports an internal dissipation mechanism rather than external shocks, which dominate the afterglow emission \citep{Fenimore99}.

Once the redshifts of the first long GRBs were determined, \citetalias{Reichart01} identified a correlation between variability $V$ and peak isotropic-equivalent luminosity $L_{\rm iso}$ for a dozen GRBs with measurable quantities: $L_{\rm iso}\propto V^{\alpha}$ with $\alpha=3.3_{-0.9}^{+1.1}$. A similar result, based on a slightly different definition of $V$, was reported by \citet{Fenimore00}, using a smaller sample. Adopting the \citetalias{Reichart01} definition of $V$, this correlation was confirmed a few years later in a sample of about 30 GRBs, despite considerable scatter \citep[hereafter G05]{Guidorzi05b} and ongoing debate about the exact value of $\alpha$ \citep{Guidorzi06b}.
 
During the {\it Neil Gehrels Swift} Observatory \citep{Gehrels04} and {\it Fermi} era, research shifted to a different temporal metric, known as the minimum variability timescale (MVT; \citealt{MacLachlan12,MacLachlan13,Golkhou14,Golkhou15}). MVT identifies the shortest timescale over which a  significant flux change occurs in an uncorrelated way, indicating a different temporal structure from the surrounding bins. It was found that MVT is anti-correlated with the bulk Lorentz factor $\Gamma$ measured from the afterglow onset time, following a relation MVT $\propto\Gamma^{-\beta}$, with $\beta$ values ranging from 2 to 4, depending on the exact definition of MVT and the GRB sample \citep{Sonbas15,Camisasca23}. Additionally, MVT was observed  to anti-correlate with $L_{\rm iso}$, after accounting for the selection effects \citep{Camisasca23}.

Interpreting the observed variability as a result of internal dissipation processes, the $e^\pm$ photosphere would smooth out all dissipation events occurring below its radius, thereby determining the observed MVT \citep{Kobayashi02,Meszaros02}. 
In a model involving a wind of shells with varying emission times and Lorentz factors $\Gamma$, smaller values of  $\Gamma$ would correspond to smaller dissipation radii and thus potentially experience stronger smoothing effects from the $e^\pm$ photosphere. This implies that slower shells would result in longer MVT. The $V$-$L_{\rm iso}$ correlation was interpreted as arising from a correlation between the jet's opening angle and the mass of the relativistic ejecta \citep{Kobayashi02}, under the assumption that the collimation-corrected gamma-ray released energy was narrowly clustered, although it was later found to be more broadly distributed than initially thought \citep{Liang08}.

Twenty years have passed since the last tests of the $V$-$L_{\rm iso}$ correlation. Today, with nearly ten times more GRBs available with suitable data for measuring both $V$ and $L_{\rm iso}$, thanks to catalogues likes {\it Swift}, {\it Fermi}, and Konus/{\it WIND}, we can re-examine this relation with greater statistical sensitivity. 
The goals of the present work are twofold: (i) to conduct an updated and more statistically sensitive test of the $V$-$L_{\rm iso}$ relation using these extensive GRB catalogues and (ii) to explore for the first time the relation between $V$ and MVT, metrics often vaguely interpreted as similar measures for variability or assumed to be strongly correlated.
 
Section~\ref{sec:data} describes the data sets, with their analysis detailed in Section~\ref{sec:ana}. The results are presented in Section~\ref{sec:res} and discussed in Section~\ref{sec:disc}, followed by the conclusions in Section~\ref{sec:conc}.
Throughout this paper, we assume the latest Planck cosmological parameters: $H_0=67.4$~km\,s$^{-1}$\,Mpc$^{-1}$, $\Omega_m=0.315$, $\Omega_\Lambda=0.685$ \citep{cosmoPlanck20}.

\section{Data sets}
\label{sec:data}
\subsection{{\it Swift}/BAT sample}
\label{sec:swift}
We selected all the long GRBs detected by {\it Swift}/BAT in burst mode from January 2005 to February 2024 with measured spectroscopic redshift and rejected all the events that had been classified as either short or short with extended emission \citep{Norris06}, whereas the long-lasting merger candidates GRB\,060614, GRB\,191019A\footnote{An alternative interpretation as a disguised tidal disruption event was also put forward for this event \citep{EylesFerris24}.}, and GRB\,211211A \citep{Gehrels06,Levan23,Yang22} were treated separately. The information on the classification of each GRB  was taken either from the BAT3 catalogue \citep{Lien16}, when available, or from the {\it Swift}/BAT team circulars. We also rejected the bursts whose LCs had not entirely been covered in burst mode, or which were affected by data gaps during the GRB.
All the bursts, for which no information was accessible on the time-integrated, {\it broad-band} spectrum apart from the one in the 15--150~keV band obtained by BAT itself, were discarded: being interested in a reliable estimate of the bolometric peak luminosity, we considered the BAT spectrum alone inadequate because of its narrow passband, which in most cases cannot constrain the peak energy of the $\nu F_\nu$ spectrum and/or the high-energy index.\footnote{We made an exception for 191019A, given the interest in this merger candidate along with the possibility of constraining the peak energy.}.

For each burst with redshift $z$, following the guidelines of the BAT team\footnote{\url{https://swift.gsfc.nasa.gov/analysis/threads/bat\_threads.html}.}, we extracted the mask-weighted LCs in the 15--150~keV passband with six different bin times, $\Delta t = 64\,(1+z)^\beta$~ms, with $\beta$ ranging from $0$ to $1$ and evenly spaced by $0.2$ increments. As it is clarified later on, this is a way to obtain all the LCs with a common bin time in the {\it comoving} frame, accounting for both cosmological time dilation and the dependence of GRB time profiles on photons' energy.
As a result, we ended up with a sample of 278~GRBs from {\it Swift}/BAT.

\subsection{{\it Fermi}/GBM sample}
\label{sec:fermi}
From the catalogue of long GRBs provided by the {\it Fermi} team, we selected all the long $(T_{90}>2$~s) GRBs from July 14, 2008 to February 4, 2024 with spectroscopically measured redshift. Long-duration merger candidates GRB\,211211A and GRB\,230307A \citep{Troja22,Gompertz23,Dichiara23,Levan24} were treated separately, as we did for the BAT sample. We also ignored the few, very bright GRBs which saturated the GBM detectors, such as GRB\,130427A and GRB\,221009A \citep{Preece14,Lesage23}. We rejected the GRBs that are affected by the simultaneous occurrence of a solar flare or whose profile was not entirely covered by the time-tagged event (TTE) mode of GBM.

For each GRB the LCs of the most illuminated NaI detectors were extracted in three different energy bands: 8--150, 150--900, and 8--900~keV energy range. Regarding the LC bin times, we adopted the same strategy as for the BAT data: six different values, $\Delta t = 64 (1+z)^\beta$~ms, where $z$ is the redshift and $\beta$ varies from 0 and 1.
Background was interpolated and subtracted using the GBM data tools\footnote{\url{https://fermi.gsfc.nasa.gov/ssc/data/analysis/gbm/gbm_data_tools/gdt-docs/}.}
\citep{GbmDataTools} following standard prescriptions (see \citealt{Maccary24} for details). For each burst we chose the GBM detectors based on the  ``scat detector mask'' entry on the HEASARC catalogue\footnote{\url{https://heasarc.gsfc.nasa.gov/db-perl/W3Browse/w3table.pl?tablehead=name\%3Dfermigbrst&Action=More+Options}}. We used the TTE data from the start of its $T_{90}$ interval to the end.

Charged particle spikes were identified and removed from the LCs as follows: whenever counts in a bin exceeded by $\ge 9\sigma$ the adjacent bins, they were tagged as due to a potential spike. If visual inspection of different GBM units confirmed the spurious nature of a possible spike by exhibiting completely different intensities, therefore incompatible with being caused by a plane electromagnetic wave, its counts were replaced with the mean of the adjacent bins.

In this way, we came up with a sample of 136~GRBs from {\it Fermi}/GBM.

\section{Data analysis}
\label{sec:ana}

\subsection{Estimate of variability}
\label{sec:Var}
For each GRB, we preliminarily determined the time window including the GRB signal, to be used for calculating the variability $V_f$; $f$ is the fraction of the total net counts, upon which the definition of variability depends in the way that is explained below. After a few attempts to find the best compromise between the need of covering the whole GRB profile and limiting the impact of noise, we opted for the following criterium: we determined the $7\sigma$ interval, whose boundaries correspond to the first and the last time bin in which the net count rate exceeds zero by $\ge 7 \sigma$, where $\sigma$ is the count rate error. This window was determined through the analysis of any given LC considering a range of bin times, from the original one to its multiples $2^n$ ($n=1,2,\ldots 7$).\footnote{This ensures that long-lasting, weak but statistically significant tails may not be cut off.} Hereafter, all the following steps will refer to the data within the $7\sigma$ window of each GRB.
To limit the impact of low-S/N GRBs, we rejected all the bursts whose total net counts have S/N$<30$.

Let $(r_i, \sigma_i)$ be the net count rate and its Gaussian error relative to the $i$-th time bin. The Gaussian limit is ensured in the case of BAT mask-weighted profiles by the central limit theorem, being the result of linear combinations of numerous independent counters, whereas the typical counts in the bin times of GBM profiles adopted in this work are always enough.

Following \citetalias{Reichart01}, variability $V_f$ is calculated as
\begin{equation}
    V_f\ =\ \frac{\sum_{i=1}^{n} [(r_i - s_{f,i})^2 - k_f\,\sigma_i^2 ]}{\sum_{i=1}^{n} (r_i^2 - \sigma_i^2)}\;,
    \label{eq:V}
\end{equation}
where $n$ is the number of bins. $\{s_{f,i}\}$ is the smoothed version of the original LC $\{r_i\}$, obtained as the convolution of $\{r_i\}$  with a boxcar window with duration $T_f$. This, in turn, is defined as the shortest cumulative time collecting a fraction $f$ of the total net counts. The interval defining $T_f$ is not necessarily contiguous, as it may be the case in the presence of quiescent times (see figures~1 and 2 of \citetalias{Reichart01}); $f$ is initially treated as a free parameter in the interval $[0.1,0.9]$.
The factor $k_f$ corrects the weight of the noise variance to be subtracted and is calculated as $(1 - 1/n_f)$, where $n_f$ is the number of bins within the smoothing boxcar window of duration $T_f$, thus equal to the rounded integer of $T_f/\Delta t$, where $\Delta t$ is the bin time of the original LC. This factor comes from the fact that, in the numerator of Eq.~(\ref{eq:V}), $r_i$ and $s_{f,i}$ are not independent. The subscript $f$ in the definition of $V$ in Eq.~(\ref{eq:V}) reminds us that it depends on $f$ through the dependence of the smoothed version of the LC on $f$.

By construction, $V_f$ ranges between 0 and 1. Both variance terms in the numerator and in the denominator of Eq.~(\ref{eq:V}) are removed of the contribution due to statistical noise ($\sigma_i^2$).

The essence of the definition of $V_f$ can be summarised in three main steps:
\begin{enumerate}
    \item determining the characteristic smoothing time $T_f$ as a function of $f$;
    \item determining the smoothed version $\{s_{f,i}\}$ of the original LC $\{r_i\}$; the subscript $f$ reminds us that the smoothed profile depends on $T_f$, which, in turn, depends on $f$;
    \item estimating how the original LC fluctuates around the smoothed version obtained in 2., removing the contribution of the noise due to counting statistics.
\end{enumerate}

To calculate $T_f$ we had to preliminarily determine the optimal bin time for any given LC: a too fine value would lead to underestimating $T_f$ because dominated by statistical fluctuations, whereas a rough resolution would cause a loss of sensitivity to the GRB temporal structures with consequent overestimate of $T_f$. The detection timescale of the narrowest, $\ge 5\sigma$ significant peak, as determined with {\sc mepsa} \citep{Guidorzi15a}, was found to be the best bin time to estimate $T_f$ in an unbiased way for all GRBs. Consequently, each LC was rebinned accordingly and then used only for the task of calculating $T_f$.

Equation~(\ref{eq:V}) is a simplified version of the original equation of \citetalias{Reichart01}: thanks to the TTE data which allow us to accumulate LCs with the desired bin time, we need not smooth the original light curve as it was the case for the binned profiles used by \citetalias{Reichart01} and \citetalias{Guidorzi05b}. We tested Equation~(\ref{eq:V}) by carrying out a suite of simulations, assuming light curves with negligible statistical uncertainties, for which $V_f$ could therefore be calculated with negligible uncertainty. We then applied Equation~(\ref{eq:V}) to a set of random noisy realisations of the same synthetic profiles and verified that we obtained unbiased and consistent values within uncertainties.

The error on $V_f$ was calculated as follows: we assumed the original LC as the set of expected rates and we generated 1000 synthetic profiles, where each simulated rate $r_{{\rm sim},i}\sim N(r_i, \sigma_i^2)$. In this way, we ended up with second-order realisations of the true (error-free and unknown) profile, so the corresponding variance of $r_{{\rm sim},i}$ is $2\sigma_i^2$, not just $\sigma_i^2$. Each synthetic profile went through the same three steps above and treated like the real LC. From the distribution of the simulated values of $V_{{\rm sim},f}$ we obtained the 90\% confidence interval, given by the 5\% and 95\% quantiles.

\begin{figure}
   \includegraphics[width=0.47\textwidth]{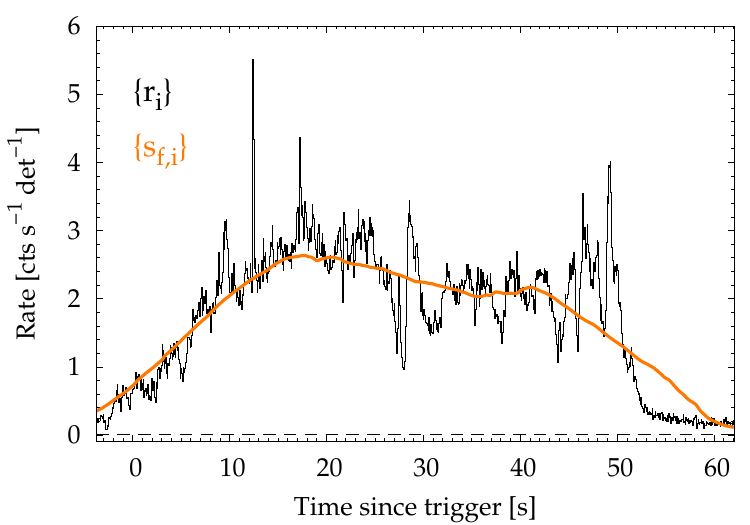}
    \caption{Illustrative example of how variability is calculated. Shown is the 15--150-keV light curve $\{r_i\}$ of GRB\,080319B as observed with BAT. The orange line shows the smoothed profile $\{s_{f,i}\}$ obtained with a smoothing timescale of $T_f=18.43$~s, which collects a fraction $f=0.45$ of the total fluence of the GRB.}
    \label{fig:080319B}
\end{figure}

Figure~\ref{fig:080319B} shows the example of the famous GRB\,080319B LC \citep{Racusin08} measured by BAT along with its smoothed profile: for that GRB, assuming $f=0.45$, it is found $T_f=18.43\pm0.13$~s.
%

\subsection{Estimate of peak luminosity}
\label{sec:Lum}
Firstly, we used estimates of $L_{\rm iso}$ derived from broad-band modeling for all BAT GRBs, for which they were already available in the literature. For BAT GRBs having both KW and GBM estimates, we systematically preferred KW \citep{KWGRBcat17, KWGRBcat21}, after making sure that they were consistent within uncertainties. The preference for KW relies in its better capability in constraining the high-energy power-law index, whereas GBM (NaI) is mainly limited by its smaller effective area above 100~keV \citep{Tsvetkova22}.
For the remaining BAT GRBs detected by KW, we adopted the same approach as in the KW catalogues: the isotropic luminosities were computed from the peak energy fluxes and $k$-corrected to the energy range $1/(1+z)$~keV--$10$~MeV.\footnote{As explained in \citet{KWGRBcat17,KWGRBcat21}, using the rest-frame upper boundary of $10\,(1+z)$~MeV instead of $10$~MeV compensates the fact that the {\it Konus}/WIND energy range extends to $10$--$25$~MeV.}
The peak spectrum was accumulated over a time interval $T_p$ around the peak, which lasted longer than 64~ms to collect enough photons. Then, under the assumption of negligible spectral evolution during $T_p$, we rescaled the average flux resulted from fitting the spectrum by the ratio $r_{64}/r_{T_p}$, where $r_{64}$ and $r_{T_p}$ are respectively the peak count rate evaluated over 64~ms and the average count rate over $T_p$. Both count rates refer to the KW net counts in the $\sim20$--$1200$~keV light curves.

For the {\it Fermi} GRBs observed from the August 4, 2008 to June 20, 2018, we used the peak luminosity provided in the {\it Fermi} spectral catalogue \citep{Poolakkil21}. For the remaining 40 more recent GBM bursts, the peak luminosity was computed as
\begin{equation}
\label{eq:lum_gbm}
    L_{\rm{iso, GBM}}\ =\ 4 \pi\ d_L^{2}\ k\ \phi_{p,8-900}\;,
\end{equation}
where $d_L$ is the luminosity distance, $k$ the k-correction and $\phi_{p, 8-900}$ the peak flux in the 8-900 keV energy range. The peak flux is computed from the best-fitting Band function of the spectrum accumulated over a $1.024$-s window centred on the peak time and modelled with the GBM data tools. We made sure that our procedure yields consistent results with the ones published in \citet{Poolakkil21} within 30\% accuracy by independently analysing a few common GRBs. For 22/40 cases, as the high-energy spectral index of the Band function was poorly constrained, we fixed it to the typical value of $-2.3$ \citep{Preece00,Kaneko06,Guidorzi11c,KWGRBcat17}. The k-correction was obtained by renormalising the peak flux in the $1/(1+z)$--$10^4/(1+z)$~keV observer-frame energy band.
For the final merged sample, for each GRB in common between KW and GBM, we opted for the $L_{\rm iso}$ estimate of the former.

Lastly, for the long-duration merger candidate GRB\,191019A \citep{Levan23} only BAT data were available. We extracted the 15-150~keV spectrum at peak centred at $0.272$~s since the trigger time with an exposure of $0.128$~s, given by the detection timescale of {\sc mepsa}. Although a simple power-law yields an acceptable fit, the photon index of $\Gamma=1.78\pm0.22$ is suggestive of the presence of the peak energy of the $\nu F_\nu$ spectrum, $E_{\rm p}$ within the BAT passband \citep{Sakamoto09}. We therefore modelled it with the Band function \citep{Band93}, fixing the low- and the high-energy indices to the typical values of $-1$ and $-2.3$, respectively. 
We found $E_{\rm p}= 62_{-20}^{+29}$~keV. Calculating the fluence in the 1--$10^4$~keV rest-frame band, the peak flux is $(1.2\pm 0.2)\times 10^{-6}$~erg~cm$^{-2}$~s$^{-1}$, corresponding to $L_{\rm iso}=(2.6\pm 0.5)\times 10^{50}$~erg~s$^{-1}$.

\section{Results}
\label{sec:res}
\begin{figure}
   \includegraphics[width=0.47\textwidth]{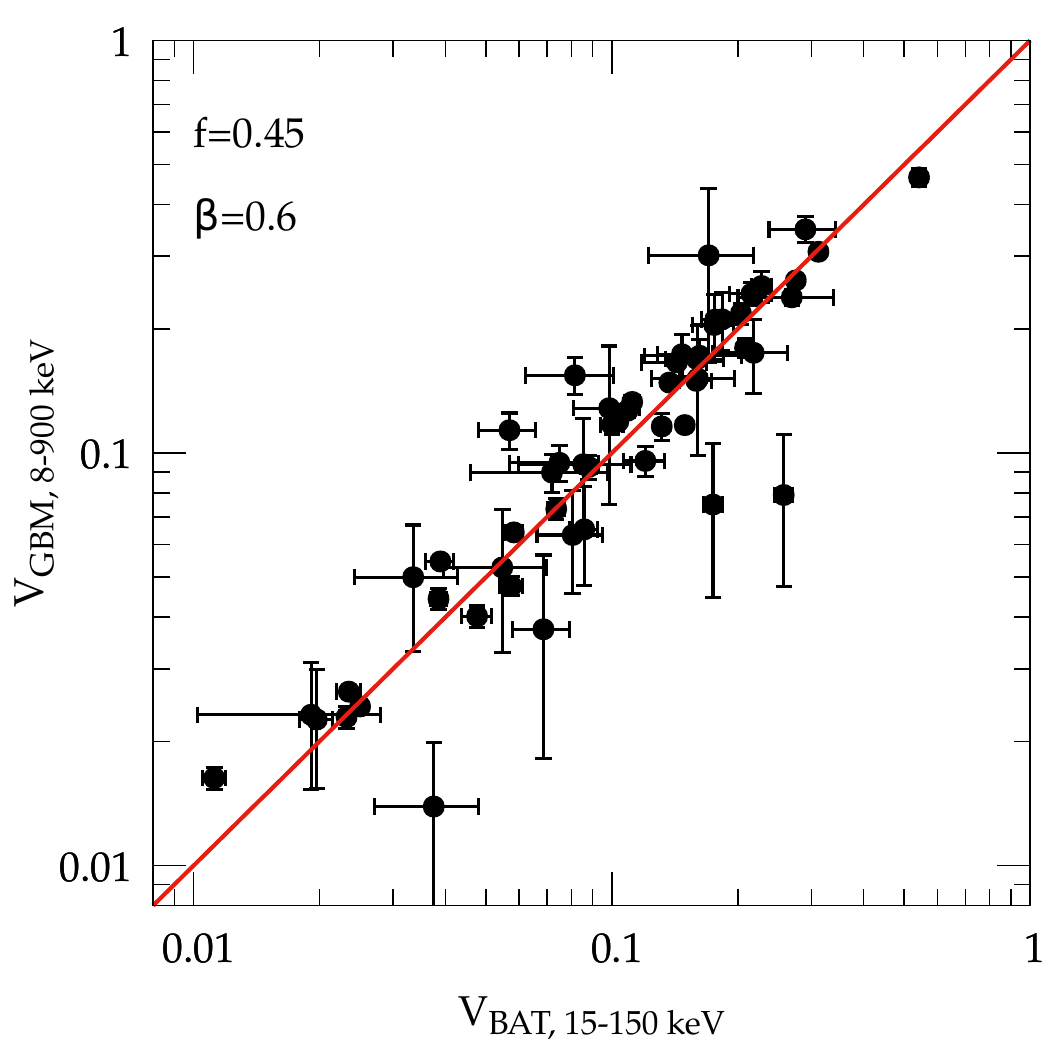}
    \caption{Comparison between estimates of variability obtained with BAT in the 15--150 keV band vs. the GBM estimates in the 8--900~keV band, obtained for a common sample and assuming $f=0.45$ and $\beta=0.6$. Equality is shown by the solid line.}
    \label{fig:V_common}
\end{figure}

From the analysis of a common sample of BAT-GBM GRBs, for which significant measures of $V_f$ were available for each of the three energy passbands (15--150, 8--150, and 150--900~keV), we found that in most cases there is a weak dependence of $V_f$ on the energy passband and on the used detector. In practice, taking the two sets of $V_f$ estimates obtained from the full passbands of each detector (15--150 vs. 8-900~keV, respectively), which conveniently have the best S/N, 85-90\% of them differ by $\lesssim 20$\%. Figure~\ref{fig:V_common} shows the comparison between the two estimates of $V_f$ for the specific case of $f=0.45$ and $\beta=0.6$. A more comprehensive comparison and analysis is reported in Appendix~\ref{sec:V_energy_dep}.
To the aim of the present investigation, a $\lesssim 20$\% mismatch in $V_f$ can be neglected, as the dynamic range of $V_f$ as well as the scatter observed in the $V_f$--$L$ plane will show. Consequently, for the majority of GRBs we may neglect the dependence of $V_f$ on the energy passbands, in agreement with what was found by \citetalias{Reichart01}.

For a common sample of 97 GRBs detected by Konus/{\it WIND} and {\it Fermi}/GBM, for which it was thus possible to independently estimate the $1$--$10^4$~keV rest-frame $L_{\rm iso}$, we compared the two sets of values.
The two sets are overall consistent over more than three decades. The distribution of the ratio $L_{\rm iso,KW}/L_{\rm iso,GBM}$ presents a median value of $1.3$ with $[1.0, 1.6]$ as interquartile range.\footnote{It is the range comprising the two central quartiles, so the 25--75 percentiles.}
That the KW estimate is $(30\pm30)$\% higher stems from its better capability in constraining the high-energy PL index, already discussed in Section~\ref{sec:Lum}. KW estimates are therefore preferable and were used for the sample of common GRBs. Overall, for the GRBs detected by GBM alone, a 30\% discrepancy has a negligible impact on the $V_f$--$L$ correlation, as it is shown in the following.

\subsection{Variability versus peak luminosity}
\label{sec:VL}
It was originally found by \citetalias{Reichart01} that the most significant correlation between $V_f$ and $L_{\rm iso}$ is obtained for $f=0.45$. $\beta$ was fixed to $0.6$, because of the contrasting effects due, on the one side, to cosmic dilation, which would demand $\beta=1$, and, on the other side, to the narrowing of pulses with energy, which would instead imply $\beta=-0.4$ (see \citetalias{Reichart01}).
\begin{figure}
   \centering
   \includegraphics[width=0.47\textwidth]{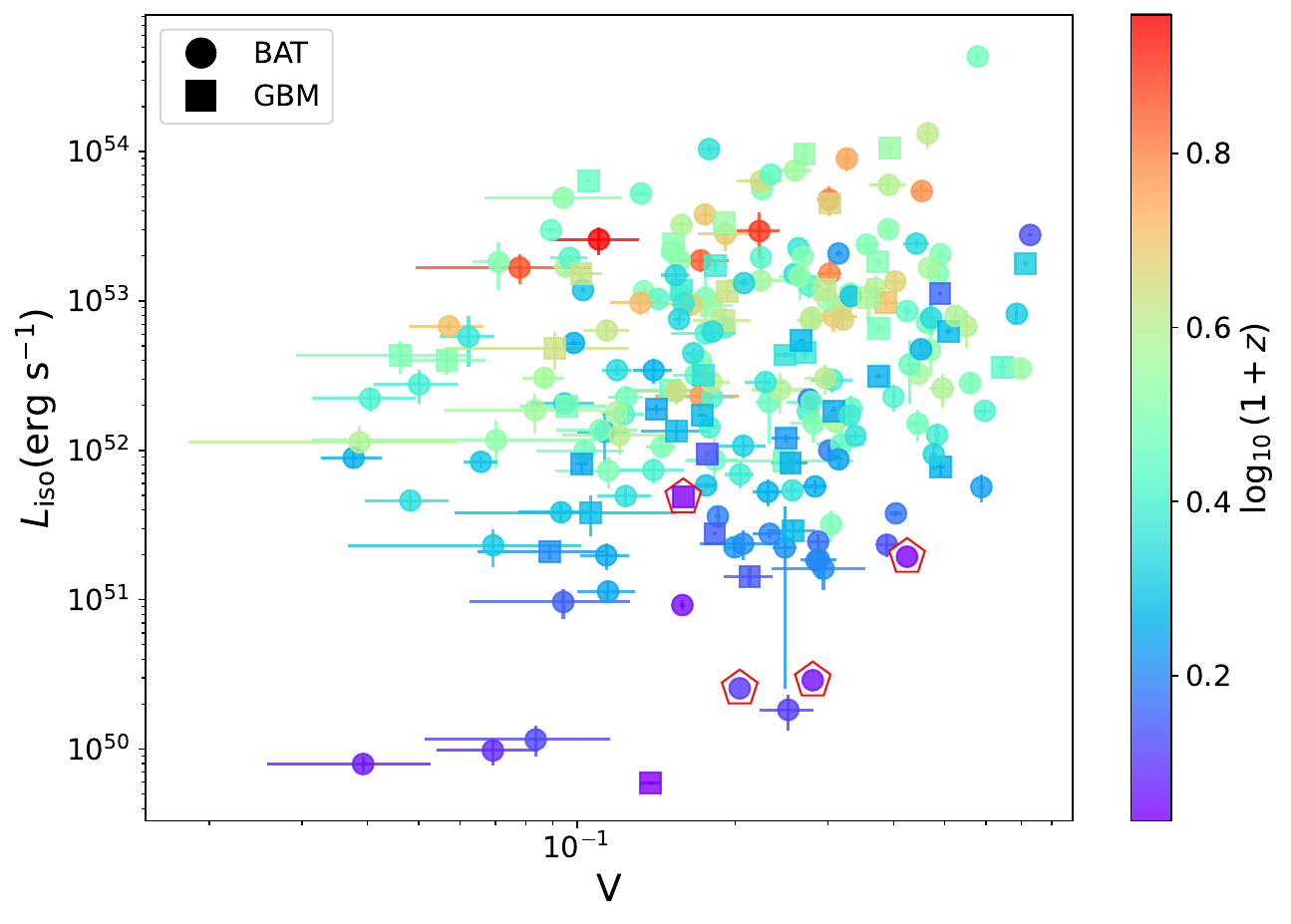}
   \caption{Variability--luminosity correlation obtained for $f=0.8$ and $\beta=0.6$. This value of $f$ gives the highest degrees of correlation. Different symbols refer to the different detectors used to measure $V_f$. For BAT-GBM shared GRBs we used BAT values. Pentagons are long-duration merger candidates. The redshift information is also available through the colour-coded scale.}
   \label{fig:VL_f0.8_beta0.6_z}
\end{figure}

First of all, in the evaluation of the statistical significance of the $V_f$--$L_{\rm iso}$ correlation that follows, we ignored the four long-duration merger candidates and in the end we show them only for comparison.

We systematically calculated the correlation coefficients (Pearson's $r$, Spearman's $\rho$, Kendall's $\tau$) between $\log{V_f}$ and $\log{L_{\rm iso}}$  for the entire grid of $(f, \beta)$ values (Section~\ref{sec:ana}).
We explored $\beta\ne 0.6$ values for the sake of completeness. For any values of $(f, \beta)$ we considered only the pairs $(V_f, L_{\rm iso})$ with a $V_f>0$ at 90\% confidence; the remaining GRBs were simply ignored, because their 90\% upper limits on $V_f$ did not turn out to be usefully constraining in the $V_f$--$L_{\rm iso}$ plane. Consequently, the number of selected GRBs varies for different $(f, \beta)$ values. Overall, the joint sample includes 216 GRBs with significant estimates for both observables: hereafter, this is referred to as the merged sample. 
%
%
\begin{table*}
\centering
\caption{Correlation coefficients of the $V_f$--$L_{\rm iso}$ relation for the joint BAT-GBM sample, assuming $\beta=0.6$.
}
\label{tab:corrcoeff}
\begin{tabular}{cccccccr}
\hline
$f$     &  Pearson's & p-value    & Spearman's & p-value    & Kendall's  & p-value    & $N_{\rm grb}^{\mathrm{(a)}}$\\
        &   $r$     & ($r$)      &   $\rho$   &  ($\rho$)  &  $\tau$    &  ($\tau$)  &\\
\hline
$0.10$  & $  -0.085$ & $    0.41$ & $  -0.075$ & $    0.46$ & $  -0.052$ & $    0.45$ & $ 99$\\
$0.15$  & $   0.055$ & $    0.55$ & $   0.021$ & $    0.82$ & $   0.012$ & $    0.84$ & $124$\\
$0.20$  & $  -0.028$ & $    0.74$ & $  -0.033$ & $     0.7$ & $  -0.023$ & $    0.68$ & $141$\\
$0.25$  & $   0.022$ & $    0.79$ & $   0.018$ & $    0.83$ & $  0.0072$ & $     0.9$ & $151$\\
$0.30$  & $    0.06$ & $    0.45$ & $   0.046$ & $    0.56$ & $   0.029$ & $    0.59$ & $158$\\
$0.35$  & $    0.16$ & $   0.037$ & $   0.097$ & $    0.21$ & $   0.066$ & $     0.2$ & $170$\\
$0.40$  & $    0.14$ & $   0.063$ & $    0.08$ & $    0.29$ & $   0.056$ & $    0.27$ & $177$\\
$0.45$  & $    0.15$ & $   0.036$ & $   0.086$ & $    0.25$ & $   0.057$ & $    0.25$ & $184$\\
$0.50$  & $     0.2$ & $  0.0052$ & $    0.13$ & $   0.066$ & $   0.087$ & $   0.073$ & $192$\\
$0.55$  & $    0.21$ & $  0.0038$ & $    0.15$ & $   0.032$ & $     0.1$ & $   0.036$ & $195$\\
$0.60$  & $     0.2$ & $  0.0048$ & $    0.16$ & $   0.022$ & $     0.1$ & $   0.028$ & $204$\\
$0.65$  & $     0.2$ & $  0.0038$ & $    0.18$ & $    0.01$ & $    0.11$ & $   0.015$ & $207$\\
$0.70$  & $    0.23$ & $ 0.00085$ & $    0.18$ & $  0.0078$ & $    0.12$ & $   0.011$ & $208$\\
$0.75$  & $    0.24$ & $ 0.00048$ & $    0.19$ & $  0.0053$ & $    0.12$ & $  0.0073$ & $210$\\
$0.80$  & $    0.25$ & $ 0.00031$ & $    0.19$ & $  0.0049$ & $    0.13$ & $  0.0057$ & $212$\\
$0.85$  & $    0.24$ & $ 0.00054$ & $    0.19$ & $  0.0051$ & $    0.13$ & $  0.0047$ & $213$\\
$0.90$  & $    0.23$ & $ 0.00085$ & $     0.2$ & $   0.004$ & $    0.13$ & $  0.0039$ & $215$\\
\hline
\end{tabular}
\begin{list}{}{}
\item[$^{\mathrm{(a)}}$]{Number of GRBs with 90\%-significant estimates of $V_f$ that were used in each case.}
\end{list}
\end{table*}

We first restrict ourselves to the $\beta=0.6$ cases for the reasons explained above. The highest degree of correlation using Pearson's $r$ is found for $f=0.8$ from a sample of 212 GRBs: the p-values associated with $r$, $\rho$, and $\tau$ are $3.1\times 10^{-4}$, $4.9\times 10^{-3}$, and $5.7\times10^{-3}$, respectively. The result is shown in Figure~\ref{fig:VL_f0.8_beta0.6_z}, whose values are reported in  Table~\ref{longtab:VL0.80_0.6}.
Alternatively, using non-parametric $\rho$ and $\tau$, the most significant case is obtained for $f=0.9$ and 215~GRBs, with p-values that are comparable with the $f=0.8$ case: $8.5\times 10^{-4}$, $4.0\times 10^{-3}$, and $3.9\times10^{-3}$, respectively for $r$, $\rho$, and $\tau$. Consequently, we may consider the $f=0.8$ and $f=0.9$ statistically equivalent in essence.

\begin{table*}
\centering
\caption{Variability and peak luminosity for the joint BAT-GBM sample obtained assuming $f=0.80$ and $\beta=0.6$ (Figure~\ref{fig:VL_f0.8_beta0.6_z}). The four GRBs in the bottom are long-duration merger candidates and were treated separately.}
\label{longtab:VL0.80_0.6}
\begin{tabular}{lcccccccr}
\hline\hline
GRB     &  $z$ & $T_f$  &  $V_f$ & $V_f$ 90\% CI$^{\mathrm{(a)}}$ & $\log{L_{\rm iso}}$ & Det$^{\mathrm{(b)}}$ &  Ref$^{\mathrm{(c)}}$ & $N_{\rm p}^{\mathrm{(d)}}$\\
        &      & (s)    &        &    &    (erg~s$^{-1}$)                        &  $V_f$   &    $L_{\rm iso}$    &\\
\hline
050219A & 0.2115  & $ 16.92 \pm  1.08 $ & $  0.069 \pm  0.015 $ & $ [ 0.044,  0.094 ] $ & $ 49.991 \pm  0.088 $ & BAT & (1) &  1\\
050315  & 1.949   & $ 48.86 \pm  2.75 $ & $  0.183 \pm  0.021 $ & $ [ 0.141,  0.210 ] $ & $ 51.937 \pm  0.165 $ & BAT & (2) &  3\\
050318  & 1.44    & $  7.74 \pm  0.33 $ & $  0.204 \pm  0.013 $ & $ [ 0.183,  0.225 ] $ & $ 51.839 \pm  0.087 $ & BAT & (2) &  3\\
050401  & 2.9     & $ 13.05 \pm  0.72 $ & $  0.370 \pm  0.015 $ & $ [ 0.343,  0.390 ] $ & $ 53.084 \pm  0.091 $ & BAT & (2) &  4\\
050820A & 2.6147  & $ 26.50 \pm  1.87 $ & $  0.224 \pm  0.031 $ & $ [ 0.177,  0.279 ] $ & $ 53.136 \pm  0.057 $ & BAT & (3) &  4\\
\hline
\end{tabular}
\begin{list}{}{}
\item[$^{\mathrm{(a)}}$]{90\% Confidence interval.}
\item[$^{\mathrm{(b)}}$]{Detector used to calculate $V_f$.}
\item[$^{\mathrm{(c)}}$]{References for $L_{\rm iso}$: (1) \citet{KWGRBcat21}; (2) \citet{Yonetoku10}; (3) \citet{KWGRBcat17}; (4) present work; (5) \citet{Poolakkil21} for GRBs before June 20, 2018; present work for later GRBs; (6) \citet{Frederiks16b};  (7) \citet{Svinkin16b}; (8) \citet{Frederiks17}; (9) \citet{Tsvetkova18}; (10) \citet{Frederiks18}; (11) \citet{Svinkin18}; (12) \citet{Frederiks18b}; (13) \citet{Frederiks18c}; (14) \citet{Tsvetkova18b}; (15) \citet{Tsvetkova19}; (16) \citet{Svinkin19};  (17) \citet{Frederiks19b}; (18) \citet{Ridnaia20b}; (19) \citet{Frederiks20b}; (20) \citet{Frederiks21}; (21) \citet{Svinkin21b}; (22) \citet{Frederiks21b}; (23) \citet{Frederiks21c}; (24) \citet{Tsvetkova22b}; (25) \citet{Frederiks23b}; (26) \citet{Svinkin24}; (27) \citet{Frederiks23c}; (28) \citet{Frederiks24}; (29) \citet{Svinkin24_240218A}; (30) \citet{Yang22}; (31) \citet{Sun23}.}
\item[$^{\mathrm{(d)}}$]{Number of peaks with S/N$>5$ taken from \citet{Guidorzi24} and \citet{Maccary24b}.}
\end{list}
\end{table*}

When we relax the constraint $\beta=0.6$, the improvement in the correlation is rather small: the smallest p-value for $r$ decreases to $1.6\times 10^{-4}$ obtained for $(f=0.75, \beta=1.0)$, whereas the smallest p-values of $\rho$ and $\tau$ become $2.2\times10^{-3}$ for $(f=0.9, \beta=0.4)$.
Therefore, admitting the possibility that $\beta$ differs from the physically grounded value of $0.6$ does not significantly improve the degree of correlation between $V_f$ and $L_{\rm iso}$. Consequently, hereafter we limit the discussion to the $\beta=0.6$ case. Table~\ref{tab:corrcoeff} reports the correlation coefficients and p-values for all values of $f$, along with the number of GRBs with significant measures that were considered in each case.

The selection of $f=0.8$ (or $f=0.9$) as the best-correlation case is the result of a multitrial process, where the optimal value was selected out of 17 trial values for $f$ in the range $0.1$--$0.9$. On the one side, these are not completely independent, being due to different ways of processing the same LCs. On the other side, they are still the results of optimal selection from multiple attempts.

To determine the effective p-value, that is the probability that the observed degree of correlation is compatible with null hypothesis of no correlation, we carried out $10^4$ simulations, each of which consisted in shuffling the array of peak luminosities and determining the best-correlation case for each synthetic sample of $(V_f, L_{\rm iso})$ pairs in the very same way as we did for the real sample. Equal or lower p-values than the corresponding lowest real ones of $(r, \rho, \tau)$ were obtained in $(26, 251, 232)$ cases: the effective p-values are, therefore, respectively $2.6\times10^{-3}$, $2.5$\%, and $2.3$\%. We conclude that the effective probability to obtain by accident an equally or more correlated sample in $V_f$--$L_{\rm iso}$ space than what shown in Fig.~\ref{fig:VL_f0.8_beta0.6_z}, under the null hypothesis of no correlation, is $\lesssim 2$\%, which is equivalent to the range $2.2$--$3.0$ $\sigma$ (Gaussian).
%

\subsection{Comparison with past results}
\label{sec:comp_past}
\begin{figure}
   \centering
   \includegraphics[width=0.47\textwidth]{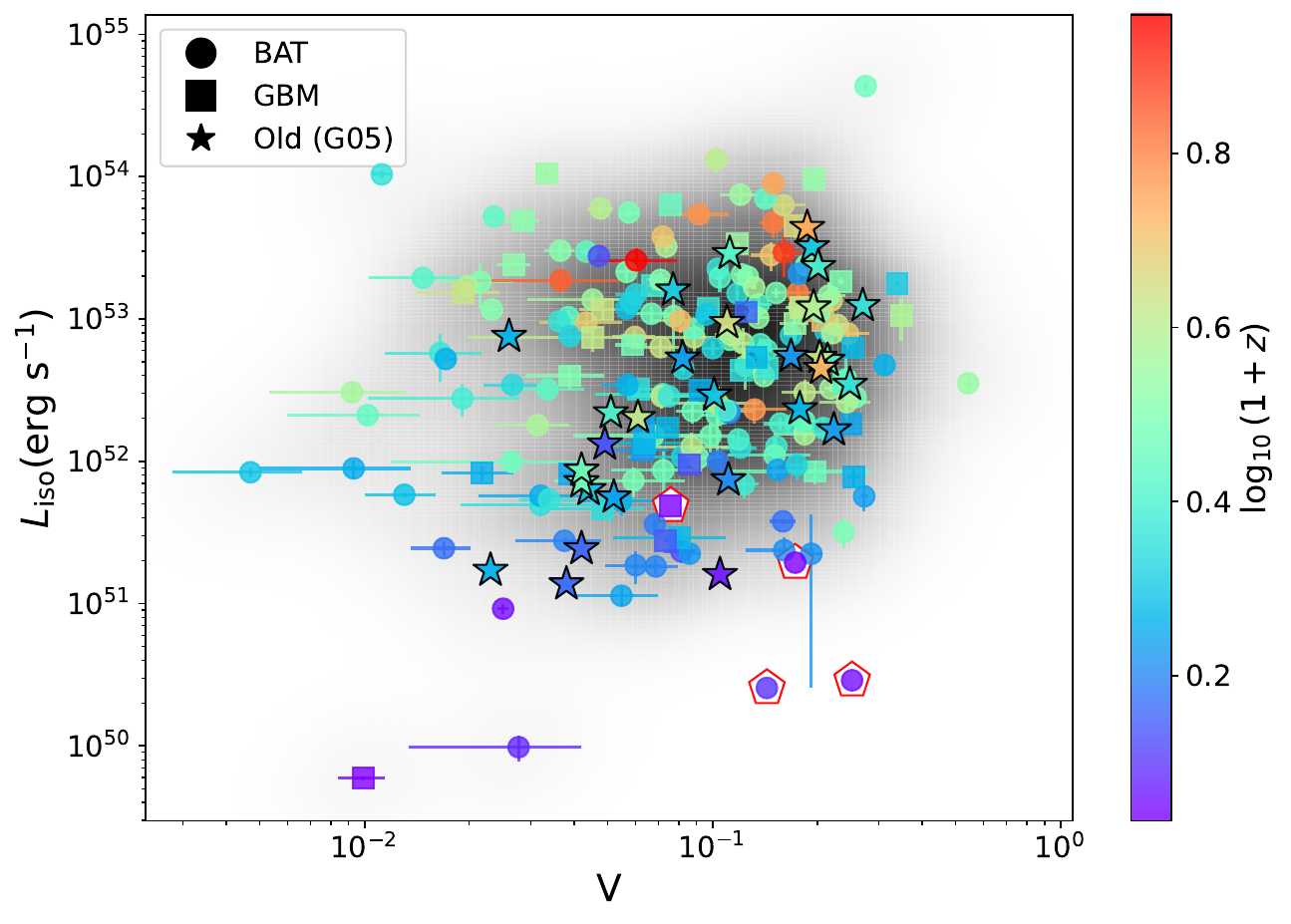}
    \caption{Variability--luminosity correlation obtained for $f=0.45$ and $\beta=0.6$. In addition to the 184 GRBs analysed in the present work, also shown (stars) are GRBs from \citetalias{Guidorzi05b}. Redshift is colour-coded. The shaded region shows a density map, obtained using a kernel density estimate, of the same data set (excluding the \citetalias{Guidorzi05b} GRBs). The four GRBs with red pentagons are long-duration merger candidates (GRB\,060614, GRB\,191019A, GRB\,211211A, and GRB\,230307A) that were considered separately.}
    \label{fig:VL_f0.45_beta0.6_z}
\end{figure}

To compare the results with what was previously obtained by \citetalias{Guidorzi05b}, we had to the use the values obtained for $f=0.45$, which was considered at the time. The resulting $V_f$--$L_{\rm iso}$ distribution is shown in Figure~\ref{fig:VL_f0.45_beta0.6_z}. Using this value for $f$, the sample of significant pairs $(V_f, L_{\rm iso})$ decreases to 184 and the p-values of the linear and non-parametric correlation tests increase to $3.6$\% and 25\%, respectively, so the previously found and very marginal evidence for correlation essentially vanishes. Table~\ref{longtab:VL0.45_0.6} reports all the values of $V_f$ and $L_{\rm iso}$ for this sample.
%

\begin{table*}
\centering
\caption{Similar to Table~\ref{longtab:VL0.80_0.6}, apart from the value $f=0.45$ (instead of $f=0.80$ of Table~\ref{longtab:VL0.80_0.6}), which was used to calculate $V_f$ and which determines the sample of GRBs with significant $V_f$ shown in Figures~\ref{fig:VL_f0.45_beta0.6_z} and \ref{fig:VL_f0.45_beta0.6_Np}.}
\label{longtab:VL0.45_0.6}
\begin{tabular}{lcccccccr}
\hline\hline
GRB     &  $z$ & $T_f$  &  $V_f$ & $V_f$ 90\% CI$^{\mathrm{(a)}}$ & $\log{L_{\rm iso}}$ & Det$^{\mathrm{(b)}}$ &  Ref$^{\mathrm{(c)}}$ & $N_{\rm p}^{\mathrm{(d)}}$\\
        &      & (s)    &        &    &    (erg~s$^{-1}$)                        &  $V_f$   &    $L_{\rm iso}$    &\\
\hline
050219A & 0.2115  & $  7.56 \pm  0.36 $ & $  0.028 \pm  0.014 $ & $ [ 0.003,  0.051 ] $ & $ 49.991 \pm  0.088 $ & BAT & (1) &  1\\
050315  & 1.949   & $ 18.67 \pm  1.10 $ & $  0.072 \pm  0.021 $ & $ [ 0.038,  0.106 ] $ & $ 51.937 \pm  0.165 $ & BAT & (2) &  3\\
050318  & 1.44    & $  3.49 \pm  0.22 $ & $  0.123 \pm  0.013 $ & $ [ 0.101,  0.144 ] $ & $ 51.839 \pm  0.087 $ & BAT & (2) &  3\\
050401  & 2.9     & $  5.08 \pm  0.72 $ & $  0.177 \pm  0.016 $ & $ [ 0.148,  0.200 ] $ & $ 53.084 \pm  0.091 $ & BAT & (2) &  4\\
050820A & 2.6147  & $  9.94 \pm  1.10 $ & $  0.045 \pm  0.016 $ & $ [ 0.018,  0.070 ] $ & $ 53.136 \pm  0.057 $ & BAT & (3) &  4\\
\hline
\end{tabular}
\begin{list}{}{}
\item[$^{\mathrm{(a)}}$]{90\% Confidence interval.}
\item[$^{\mathrm{(b)}}$]{Detector used to calculate $V_f$.}
\item[$^{\mathrm{(c)}}$]{References for $L_{\rm iso}$: (1) \citet{KWGRBcat21}; (2) \citet{Yonetoku10}; (3) \citet{KWGRBcat17}; (4) present work; (5) \citet{Poolakkil21} for GRBs before June 20, 2018; present work for later GRBs; (6) \citet{Frederiks16b};  (7) \citet{Svinkin16b}; (8) \citet{Frederiks17}; (9) \citet{Tsvetkova18}; (10) \citet{Frederiks18}; (11) \citet{Svinkin18}; (12) \citet{Frederiks18b}; (13) \citet{Frederiks18c}; (14) \citet{Tsvetkova18b}; (15) \citet{Tsvetkova19}; (16) \citet{Svinkin19};  (17) \citet{Frederiks19b}; (18) \citet{Ridnaia20b}; (19) \citet{Frederiks20b}; (20) \citet{Frederiks21}; (21) \citet{Svinkin21b}; (22) \citet{Frederiks21b}; (23) \citet{Frederiks21c}; (24) \citet{Tsvetkova22b}; (25) \citet{Frederiks23b}; (26) \citet{Svinkin24}; (27) \citet{Frederiks23c}; (28) \citet{Frederiks24}; (29) \citet{Svinkin24_240218A}; (30) \citet{Yang22}; (31) \citet{Sun23}.}
\item[$^{\mathrm{(d)}}$]{Number of peaks with S/N$>5$ taken from \citet{Guidorzi24} and \citet{Maccary24b}.}
\end{list}
\end{table*}

The GRBs considered in \citetalias{Guidorzi05b} are 32: ignoring GRB\,980425, which is a peculiar low-luminosity event \citep{Kulkarni98,LiChevalier99,Pian00,Soderberg04,Ghisellini06}, GRB\,050603 whose redshift was later questioned \citep{Hjorth12}, and other 5 {\it Swift}/BAT GRBs which are already included in the present sample, we are left with 25 additional GRBs from \citetalias{Guidorzi05b}, adding which we end up with a sample of 209 GRBs, whose p-values are $4.7\times 10^{-3}$ (Pearson's) and $3.8$ and $4.1$\% for the other non-parametric tests. The luminosity values in \citetalias{Guidorzi05b} were calculated in the 100-1000~keV band, as originally done by \citetalias{Reichart01}. We then replaced their luminosity values with the broadband $1$--$10^4$~keV rest-frame analogues as reported in the literature as derived from broad-band spectroscopy. Compared with the \citetalias{Guidorzi05b} values, luminosities increased in most cases by a factor between 2 and 3.

In Appendix~\ref{app:compat_G05} we estimated a $\gtrsim 3$\%  probability that the correlation assessed in \citetalias{Guidorzi05b} was accidental and caused by the poor sampling of the variability-luminosity space. 
On top of that, it is also possible that different selection effects between the joint sample of BAT-GBM of the present work and the early one of \citetalias{Guidorzi05b} also play a role: in particular, both GRB samples inevitably depend on the different suites of prompt optical followup facilities that enabled the afterglow identification and secured the redshift measurement.

\subsection{Relation with the number of peaks}
\label{sec:Np}
\begin{figure}
   \centering
   \includegraphics[width=0.47\textwidth]{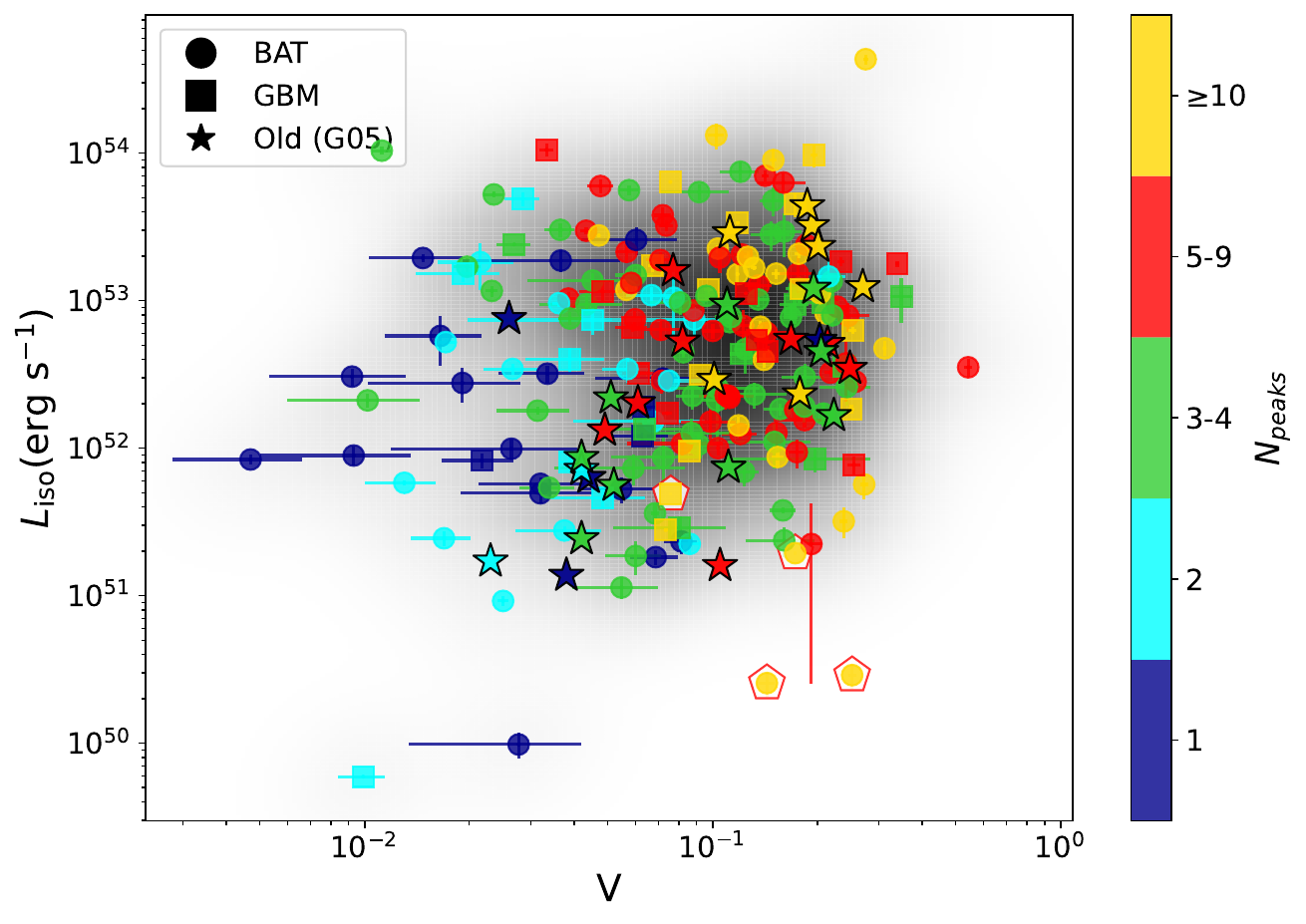}
   \caption{Same plot as that of Figure~\ref{fig:VL_f0.45_beta0.6_z}, except for the colour-code, which corresponds to five different classes of number of peaks.}
    \label{fig:VL_f0.45_beta0.6_Np}
\end{figure}

Figure~\ref{fig:VL_f0.45_beta0.6_Np} is the same as Figure~\ref{fig:VL_f0.45_beta0.6_z}, except for the colours which display the number of peaks of each GRB, $N_{\rm peaks}$, as it was determined by us in \citet{Guidorzi24} and in \citet{Maccary24b} using {\sc mepsa} \citep{Guidorzi15a} and selecting peaks with S/N$>5$.

\begin{figure*}
   \centering
   \includegraphics[width=0.9\textwidth]{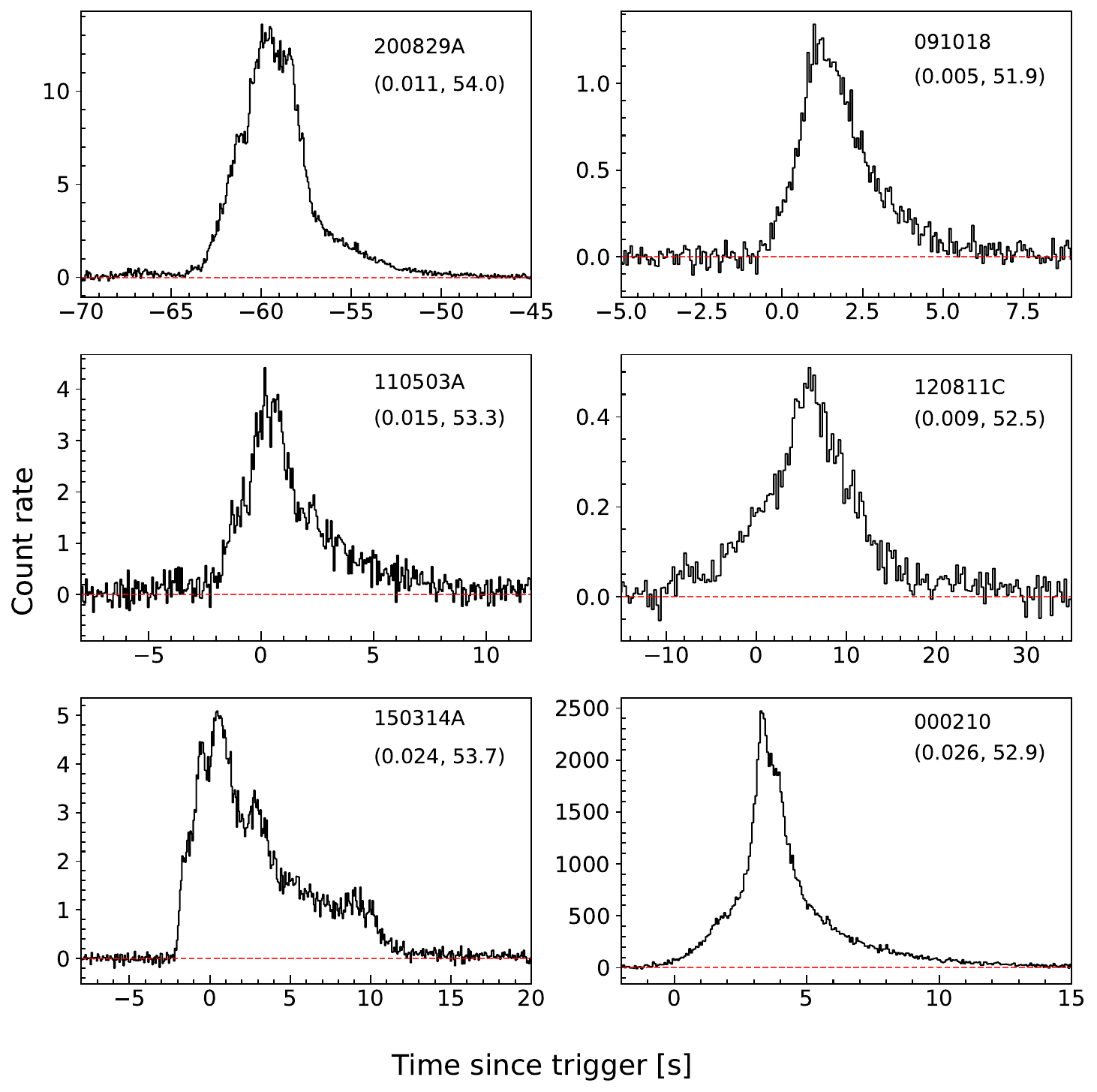}
   \caption{Collection of 6 observer-frame GRBs with medium/high luminosity and low variability. Each panel reports the GRB name along with the $(V_f, \log{L_{\rm iso}})$ pair. All of them are BAT bursts (15--150~keV), except for {\it BeppoSAX} GRB\,000210 (40--700~keV) in the bottom right panel. In the case of BAT LCs, count rates are expressed in count~s$^{-1}$ per fully illuminated detector for an equivalent on-axis source.}
    \label{fig:singlepulse}
\end{figure*}

The additional information supplied by $N_{\rm peaks}$ helps to characterise the GRBs with low $V_f$ that were mostly missing in the early sample of \citetalias{Guidorzi05b} and which contributed to demote the correlation.
Most GRBs with $V_f<0.05$ have very few peaks (mainly 1 or 2) and are 25\% of the whole sample. On average, these low-$V_f$ GRBs exhibit comparable luminosities with the complementary sample of more variable GRBs, except for a couple of them with $L_{\rm iso}\lesssim 10^{50}$~erg~s$^{-1}$.
Figure~\ref{fig:singlepulse} displays the LCs of five low-$V_f$ GRBs, along with \citetalias{Guidorzi05b} GRB\,000210\footnote{This {\it BeppoSAX} gamma-ray luminous and optically dark GRB is discussed in detail in \citet{Piro02}.}: all of them have $V_f\lesssim0.03$, in line with a relatively smooth profile, and $L_{\rm iso}$ in the range $10^{52}$--$10^{54}$~erg~s$^{-1}$.
Apart from some modulation in GRB\,150314A, their profiles look like smooth, fast-rise exponential-decay (FRED) pulses: in principle, an external origin for their prompt emission cannot be excluded (a thorough analysis of how tenable this scenario is for each of them, taking into account the corresponding afterglow multi-band data sets, is beyond the scope of the present work). Specifically, in the case of GRB\,200829A an external origin has been put forward \citep{Li23c}, whereas \citet{Samuelsson22} discuss a photospheric origin for GRB\,150314A. 
%

\subsection{Long-duration merger candidates}
\label{sec:longmerger}
Among the GRBs detected by either BAT or GBM, there are at least four credible long-duration merger candidates known to date: GRB\,060614 \citep{DellaValle06,Fynbo06}, GRB\,191019A \citep{Levan23}, GRB\,211211A \citep{Yang22,Troja22}, and GRB\,230307A \citep{Levan24}. Actually, one should also consider GRB\,060505 \citep{Fynbo06}: however, we ignored it because of the low S/N of BAT data along with its controversial nature \citep{McBreen08}.
Although they were not included in the samples for the statistical analysis of the $V_f$--$L_{\rm iso}$ correlation, we added them in Figures~\ref{fig:VL_f0.45_beta0.6_z} and \ref{fig:VL_f0.45_beta0.6_Np}.
\begin{figure*}
   \sidecaption
   \includegraphics[width=0.6\textwidth]{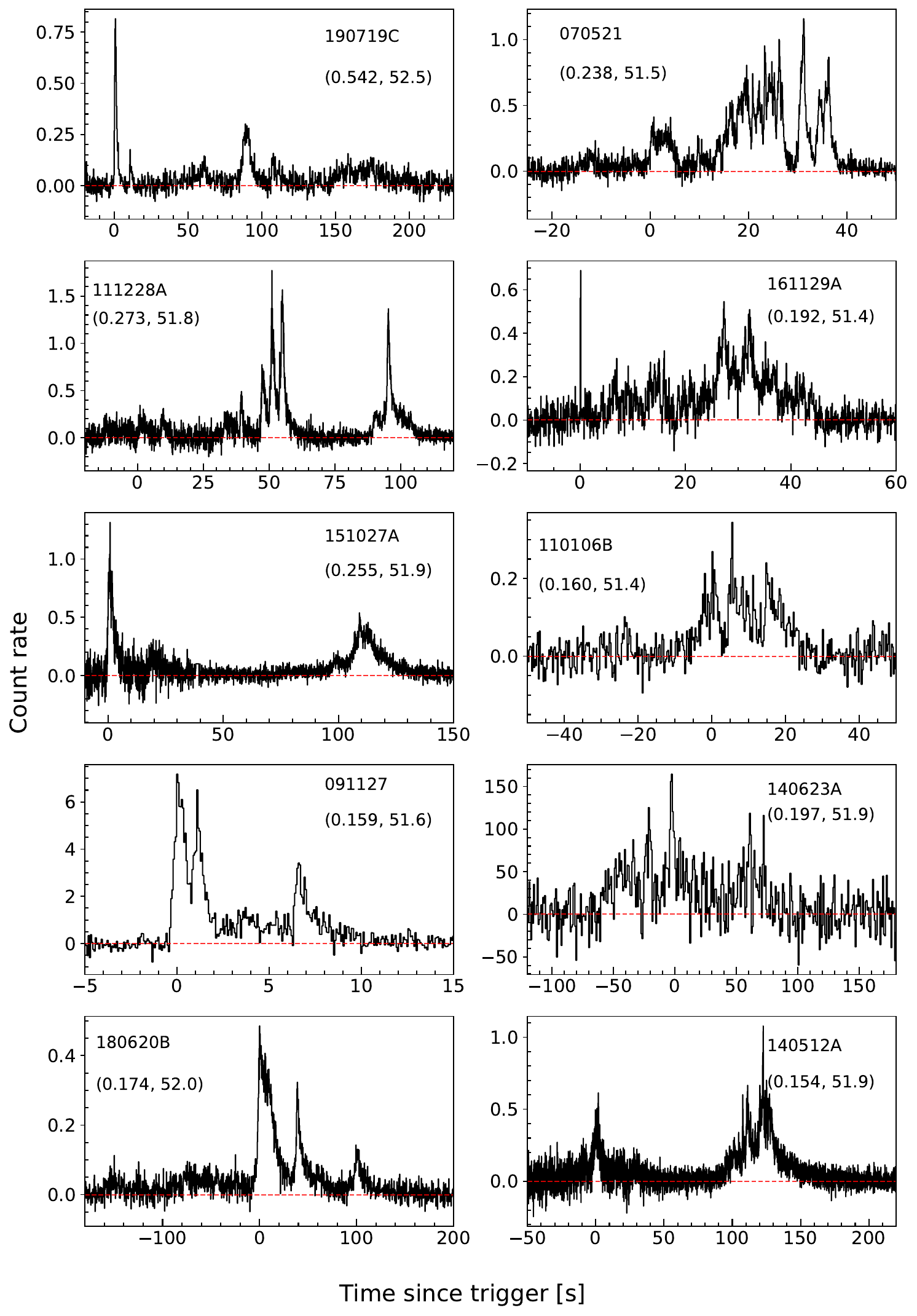}
   \caption{Collection of 10 observer-frame GRBs with high variability, $V_f>0.1$, and relatively low luminosity, $L_{\rm iso}\lesssim 10^{52}$~erg~s$^{-1}$. Each panel reports the GRB name along with the $(V_f, \log{L_{\rm iso}})$ pair. All of them are BAT bursts (15--150~keV), except for {\it Fermi} GRB\,140623A (8--900~keV). In the case of BAT LCs, count rates are expressed in count~s$^{-1}$ per fully illuminated detector for an equivalent on-axis source.}
    \label{fig:highVlowL}
\end{figure*}

Two of them (GRB\,060614 and GRB\,191019A) lie within a region of their own: $V_f>0.1$ and $L_{\rm iso}<10^{51}$~erg~s$^{-1}$. GRB\,211211A has comparable $V_f$, but higher $L_{\rm iso}$ ($2\times10^{51}$~erg~s$^{-1}$), still lying in the outskirts of the distribution of standard long GRBs. Only GRB\,230307A lies within a more densely populated region. All of them feature many peaks (ranging from 14 to 54), which contributed to the large values of $V_f$. That three out of four known candidates lie off the population distribution in the $V_f$--$L_{\rm iso}$ plane, suggests that the unusual combination of $V_f\gtrsim 0.1$ and $L_{\rm iso}\lesssim 10^{51}$~erg~s$^{-1}$ might be a good indicator for a merger origin. In this respect, a possible closely related indicator was already tentatively identified in the MVT \citep{Camisasca23,Camisasca23b,Veres23}.

To see whether there are other similar and as-yet disguised merger candidates, we explored the LC morphology of the GRBs that lie mostly in the high-$V_f$/low $L_{\rm iso}$. Figure~\ref{fig:highVlowL} displays a collection of 10 such GRBs, all of which share the following properties: $V_f>0.1$ and $L_{\rm iso}\lesssim 10^{52}$~erg~s$^{-1}$~s, with the only exception of 190719C ($L_{\rm iso}=3\times 10^{52}$~erg~s$^{-1}$) which has the highest $V_f$ of all GRBs and whose projected offset seems to be more typical of a merger event and rather large for a collapsar event, although not unprecedented \citep{Rossi19}.
Among these 10 GRBs, there is a couple of them for which the collapsar identity was firmly established by the evidence for an associated SN: GRB\,091127 \citep{Cobb10,Berger11} and GRB\,111228A \citep{Klose19}. Yet, there are interesting events which look like short GRB with extended emission, such as GRB\,161129A, as it was also noted by the {\it Swift} team: however, its nature remained inconclusive, given the spectral softness of the initial spike in comparison with the bulk population of short GRBs with and without extended emission \citep{Barthelmy16}.

\subsection{Variability and minimum variability timescale}
\label{sec:MTV_vs_V}
\begin{figure}
   \centering
   \includegraphics[width=0.47\textwidth]{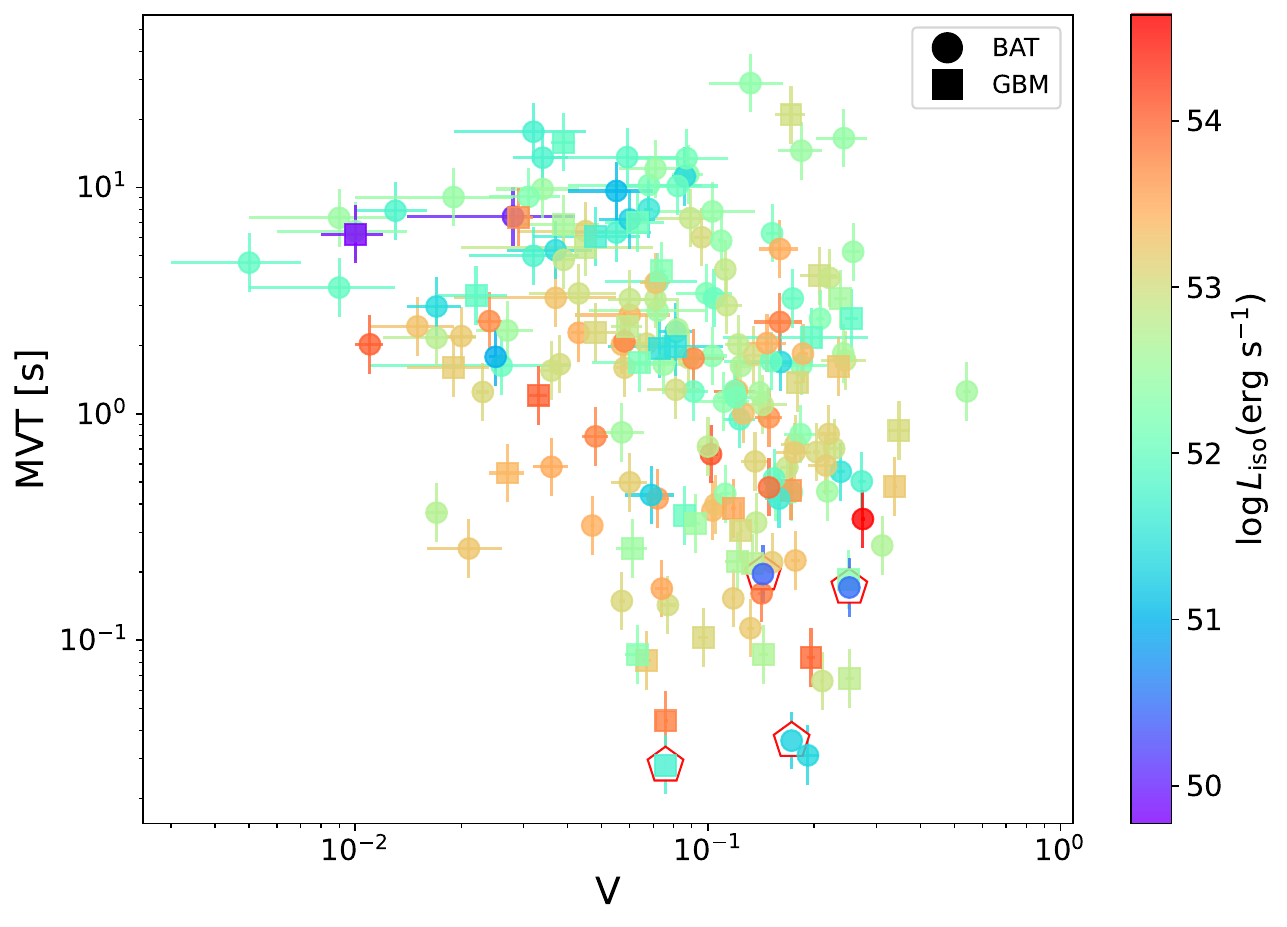}
   \caption{MVT versus variability for a subsample of 184 GRBs. Pentagons are the four long-duration merger candidates. MVT values were calculated as in \citet{Camisasca23}. Colour-coded is $L_{\rm iso}$.}
    \label{fig:MTV_vs_V}
\end{figure}

Given that GRB time variability is a recurring general property that is often called for and interpreted in the literature, it is worth investigating and clarifying the relation between the definition of variability $V_f$ considered in this work and the concept of MVT. 
To better illustrate this point, Figure~\ref{fig:MTV_vs_V} shows the two quantities calculated for a subsample of 184 GRBs.
MVT values were either taken from or calculated as in \citet{Camisasca23}.
The two quantities are clearly neither independent nor tightly correlated. In fact, MVT identifies the shortest timescale over which a significant and uncorrelated flux variation is observed, irrespective of the overall properties of the whole LC, whereas $V_f$ quantifies how much variance lies at short timescales with respect to long ones, where the short/long separation is dictated by the net time interval over which a sizeable fraction of energy is released (Section~\ref{sec:Var}).

It is, therefore, no wonder that for $V_f\gtrsim 0.1$ all possible MVT values from $0.01$ to $\sim30$~s are observed: a high value of $V_f$ does not necessarily imply the presence of narrow pulses, but it could also be obtained for a GRB featuring several broad peaks with different timescales and interspersed with quiescent times. Instead, most GRBs with $V_f\lesssim$ few $\times 10^{-2}$ are inevitably missing narrow spikes, such that their MVT is $\gtrsim$ a few seconds.

\section{Discussion}
\label{sec:disc}

Variable GRBs should arise when internal dissipation processes (e.g., internal shocks) occur outside the  $e^\pm$ pair photosphere, while X-ray rich bursts may arise from the processes occurring below it \citep{Kobayashi02,Meszaros02}. Let $d$ and $D$ be the width and separation of the random shells which characterise a GRB jet with velocity irregularity.  The hydrodynamic timescale $d/c$ and the angular spreading timescale $D/c$ determine the rise and the decay time of a gamma-ray pulse, respectively. Since most observed pulses rise more quickly than they decay, the pulse width and pulse separation are mainly determined by the angular spreading time $D/c$ \citep{Norris96,RydePetrosian02}. 

The distribution of peak separations\footnote{Also known as waiting times.} $\{D/c\}$ for a GRB light curve usually has a large dispersion \citep{NakarPiran02a,Guidorzi15b}. While the fastest variability timescale in a GRB can be as short as a few tens of milliseconds \citep{Camisasca23}, the largest peak separation or the total duration is usually much longer. Since shell collisions producing narrow pulses typically occur at small radii $\sim$ $D \Gamma^2$, the photosphere might obscure these, leaving primarily the wider pulses visible. This will make the temporal profile smooth. If a jet has a smaller typical Lorentz factor, a larger fraction of the collisions occur at small radii $\sim D\Gamma^2$ below the photosphere; therefore, the smoothing effect is expected to be stronger. 

The observed correlation between luminosity and variability in GRBs has been interpreted as reflecting a relation between the GRB jet's opening angle and the mass involved in the explosion \citep{Kobayashi02}. Since the beaming-corrected gamma-ray energy was believed to be narrowly clustered in the pre-{\it Swift} era, narrower jets would emit brighter gamma-ray emission. If narrower jets have typically higher Lorentz factors due to smaller mass loading, the photosphere could induce the luminosity and variability correlation. For the increased number of GRBs observed in the {\it Swift} era, it was found that the distribution of beaming-corrected gamma-ray energy is broader than previously thought, although still narrower than the distribution of isotropic energy \citep{Liang08}. 

The correlation between luminosity and variability is not significant for our larger sample of GRBs. To investigate potential indications of the photospheric effect in the GRB light curves, we test for a correlation between Lorentz factors and variability  $V_f$, as slower jets might induce a stronger photospheric smoothing effect.
However, we again find no significant correlations for the 37 GRBs, whose Lorentz factors were estimated from the afterglow onset times $t_p$ \citep{Ghirlanda18}. The p-values associated with linear and non-parametric correlation coefficients are 4 and 24\%, respectively. This contrasts with MVT, which shows a negative correlation with the Lorentz factor, albeit with significant scatter: MVT $\propto \Gamma^{-2}$ (see fig.~13 of \citealt{Camisasca23}). This relation is consistent with the photospheric model in which  MVT is determined by the curvature timescale at the photoshere: MVT $\sim R_{\pm}/c\Gamma^2$, although the radii estimated from MVT and $\Gamma$ are rather large, $10^{15-17}$ cm, for the photospheric radii $R_\pm$\citep{Camisasca23,Kobayashi02}.

As shown in Fig.~\ref{fig:MTV_vs_V} and discussed in Section~\ref{sec:MTV_vs_V}, the correlation between MVT and the variability $V_f$ is very weak. The average power density spectrum (PDS) of very long GRBs (with $T_{90} >100$s) is known to follow a power-law distribution characterised by a slope $\alpha \sim -5/3$, featuring a clear break at about 1 Hz \citep{Beloborodov00a,Guidorzi12,Dichiara13a}. Shorter GRBs exhibit PDS slopes that are more significantly influenced by statistical fluctuations. The lack of correlation between MVT and variability $V_f$ can be attributed to the dominance of significant components in the light curve power spectrum at timescales substantially longer than the MVT (i.e., $V_f$ is also determined by the pulses with longer timescales as well as by quiescent times). Considering MVT is better correlated with $L_{\rm iso}$ and $\Gamma$, it is likely to be more sensitive to the photospheric cut-off of the variability timescales; or, intrinsically the absolute measure of the short timescale with significant variance is correlated with them. If MVT is set by the photospheric effect, then the 1 Hz break in the averaged PDS should be due to the intrinsic nature of the central engine activity (i.e. the intrinsic separation distribution \{D\}).

A possible factor which weakens the correlation between Lorentz factors and variability $V_f$ can be the separation distribution $\{D\}$ which can vary significantly between events intrinsically. Even if the average PDS follows the power-law, especially for events with a small number of peaks, the statistical fluctuations can be significant. 
If some events intrinsically lack narrowest components or longer-timescale components in the power spectrum dominate, their variability would be insensitive to the jet’s Lorentz factor and is always low. Another potential weakening factor is the errors in the estimates of Lorentz factors. Numerous collisions happen during an evolution of multiple shells. Each collision produces a pulse. However, the main pulses are produced by collisions between the fastest shells, $\sim\Gamma_{\rm max}$, and the slowest shells, $\sim\Gamma_{\rm min}$. Such a collision happens at $R\sim \Gamma_{\rm min}^2 D_i$ for an initial separation $D_i$. The suppression of narrow peaks and consequent smoothing effect basically depend on how much the photosphere radius $R_{\pm}$ is larger than $\Gamma_{\rm min}^2 D_{\rm min}$. 
The Lorentz factor based on the afterglow onset time gives a characteristic value after the internal dissipation process is settled, but it might not correlate with the minimal value $\Gamma_{\rm min}$ of the intrinsic distribution well. 

In addition, it is also possible that the presence of GRBs, whose prompt emission has a completely different origin, contributed to weaken the possible correlation between $V_f$ and $L_{\rm iso}$. Specifically, this could be the case of low-$V_f$ GRBs, having just one broad peak and with a typical $L_{\rm iso}$: their prompt emission might have an external origin, marking the high-energy afterglow onset (Section~\ref{sec:Np} and Fig.~\ref{fig:VL_f0.45_beta0.6_Np}). For five of the six external-shock GRB candidates shown in Fig.~\ref{fig:singlepulse}, the upper limits of the afterglow onset time $t_p$ are available (see Table~\ref{tab:EScases}). We find tight upper limits especially for four of these events. The afterglow onset could happen right after/during the prompt gamma-ray emission, or the prompt gamma-ray emission itself could be the afterglow onset as we propose the external shock origin. Interestingly, two of these events (GRB\,091018 and GRB\,120811C) have very low $E_p$: taking the $E_p$ distribution of the merged sample of 317 GRBs from \citet{KWGRBcat17,KWGRBcat21}, only 1 ($\sim$$0.3$\%) and 33 ($10$\%) have equal or smaller $E_p$ than those two bursts, respectively.

%
\begin{table}
\caption{Upper limits on the observed afterglow onset time $t_p$, prompt duration $T_{90}$, and peak energy $E_p$ of the time-integrated $\nu F_\nu$ spectrum for 5 external-shock GRB candidates shown in Fig.~\ref{fig:singlepulse}.} 
\label{tab:EScases} 
\centering 
    \begin{tabular}{lrrr}
    \hline\hline
    GRB      & $T_{90}$ (s) & $t_p$ (s) & $E_p$ (keV)\\ 
    \hline
    091018      & $4.4$     & $< 138$   & $27_{-4}^{+2}$\\ 
    110503A     & $58.7$    & $< 275$   & $220\pm 12$\\
    120811C     & $24.3$    & $ < 1020$ & $49\pm 3$\\
    150314A     & $14.8$    & $< 135$   & $350\pm 10$\\
    200829A     & $13.0$    & $< 100$   & $336\pm 11$\\
    \hline 
    \end{tabular}
    \tablefoot{Upper limits on the afterglow onset times are from \citet{Ghirlanda18} and \citet{Li23c}. Peak energies are from \citet{KWGRBcat17,KWGRBcat21,Ridnaia20b}. All values refer to the observer frame.}
\end{table}

Finally, in Section~\ref{sec:longmerger} and in Figure~\ref{fig:VL_f0.45_beta0.6_z} we showed that three out of four known long-duration compact binary merger candidates lie off the bulk of long GRBs, in the region with $V_f>0.1$ and $L_{\rm iso}< 10^{51}$~erg~s$^{-1}$, with the fourth, GRB\,230307A, lying closer to other typical long GRBs. Additionally, considering their MVT (Fig.~\ref{fig:MTV_vs_V} and \citealt{Camisasca23}), these events appear to be  characterised by the rare combination of a very small MVT (a few ten ms), high variability, and relatively low luminosity. We tentatively identified other interesting events, displayed in Fig.~\ref{fig:highVlowL}, with analogous features: although for a couple of them an associated SN was found, the remaining candidates might be worth a deeper investigation.

\section{Conclusions}
\label{sec:conc}

Twenty years after the launch of {\it Swift} and its subsequent enhancement by {\it Fermi}, the number of GRBs with measured redshifts has significantly increased. This growth necessitates a new, more statistically robust examination of the variability-luminosity relation. Previously, this relation was reported based on a sample of approximately 30 GRBs available a few years after the initial afterglow discoveries.

The aim of this study is testing the correlation by utilising the extensive data sets available today. Using a sample of $216$ GRBs detected by {\it Swift, Fermi}, and {\it Konus/WIND}, each with robust estimates for both variability $V_f$ and isotropic-equivalent peak luminosity $L_{\rm iso}$, we found that the scatter has increased to such an extent that the correlation can no longer be considered statistically significant (p-value $\lesssim 2$\%).

The definition of $V_f$ adopted in this study, originally provided by \citetalias{Reichart01}, measures the temporal power of short-to-intermediate timescales with respect to the total power, which includes the contribution from all timescales.
This definition of $V_f$ is not be confused with that of MVT \citep{MacLachlan13,Golkhou14,Golkhou15,Camisasca23}, another observable used to characterise GRB variability. MVT corresponds to the shortest timescale over which a statistically significant and uncorrelated flux change is observed, regardless of the power in long timescales.
Here, we investigated for the first time the relation between $V_f$ and MVT and found a weak correlation (Fig.~\ref{fig:MTV_vs_V}). This finding helps explain why MVT is found to anti-correlate with $L_{\rm iso}$ and the Lorentz factor $\Gamma$ (measured from the afterglow onset time), 
despite significant scatter \citep{Camisasca23}, while $V_f$ does not.

When internal dissipation occurs within the $e^{\pm}$ photosphere, the resulting gamma-ray light curve is smoothed out. Narrower pulses, produced at smaller radii, are particularly sensitive to this photospheric cut-off. 
The stronger correlation of $L_{\rm iso}$ and $\Gamma$ with MVT compared to $V_f$ suggests that $V_f$ is mainly influenced by pulses and quiescent periods with timescales much longer than the MVT scale. MVT might be determined by the photospheric effect or it is intrinsically correlated with $L_{\rm iso}$ and $\Gamma$ due to the unknown nature of the central engine. 

Furthermore, we identified several GRBs with a single broad and smooth peak, low $V_f$ and typical $L_{\rm iso}$, whose origin may be attributed to
external shocks, differing from the majority of the observed GRBs (Fig.~\ref{fig:singlepulse}). This scenario is supported by the tight upper limits on the afterglow onset times determined from early optical afterglow observations. Notably, the prompt emissions of two of them (GRB\,091018 and GRB\,120811C) are very soft, having peak energies of the time-average $\nu\,F_\nu$ spectrum in the low tail of the observed population.

Lastly, the combination of high variability ($V_f>0.1$), relatively low luminosity ($L_{\rm iso}<10^{51}$~erg~s$^{-1}$), and short MVT ($\lesssim 0.1$~s; \citealt{Camisasca23}) appears to be a promising indicator for a compact binary merger origin, despite the long duration and deceptive time profile. We have tentatively identified other potential candidates with similar characteristics.

\begin{acknowledgements}
D.F., A.L., A.R., D.S., M.U. acknowledge financial support from the basic funding program of the Ioffe Institute FFUG-2024-0002. A.T. acknowledges financial support from ``ASI-INAF Accordo Attuativo HERMES Pathinder operazioni n. 2022-25-HH.0'' (discussion of the results, editing the paper) and the basic funding program of the Ioffe Institute FFUG-2024-0002 (computation of $L_\textrm{iso})$. L.F. acknowledges support from the AHEAD-2020 Project grant agreement 871158 of the European Union’s Horizon 2020 Program. We thank the reviewer for their detailed and constructive report.
\end{acknowledgements}


\begin{appendix}

\section{Dependence of variability on energy passband}
\label{sec:V_energy_dep}
\begin{figure}
   \includegraphics[width=0.47\textwidth]{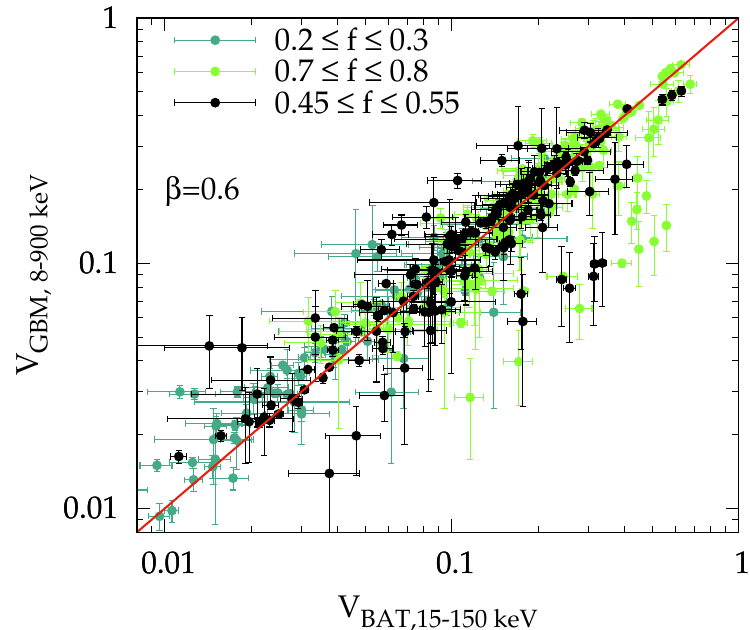}
    \caption{Comparison between estimates of variability obtained with BAT in the 15--150 keV band vs. the GBM estimates in the 8--900~keV band, obtained for a common sample with significant measures of variability. Three different ranges for $f$ are shown.}
    \label{fig:V_common2}
\end{figure}

The analysis of a common sample of BAT-GBM GRBs having significant measures of $V_f$ for each of the three energy passbands (15--150, 8--150, and 150--900~keV) showed that in most cases there is a weak dependence of $V_f$ on the energy passband. Figure~\ref{fig:V_common2} shows the comparison of $V_f$ obtained from the total-passband light curves of the two detectors. Unlike Figure~\ref{fig:V_common}, which is limited to the $f=0.45$ case, Fig.~\ref{fig:V_common2} shows all values of $f$ (see Section~\ref{sec:Var} for a definition of $f$).
\begin{figure}
   \includegraphics[width=0.47\textwidth]{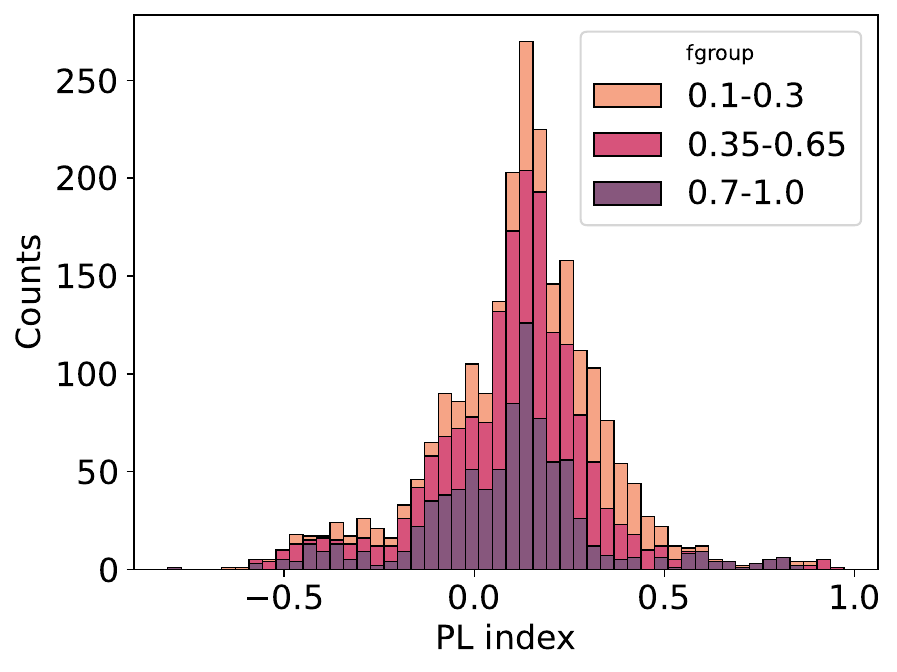}
    \caption{Distributions of the PL index $\alpha$ that models the dependence of variability on photon energy as $V\propto E^{\alpha}$ for a sample of 44 GRB shared by BAT and GBM having significant measures of $V_f$ for all of the three energy passbands (15--150, 8--150, and 150--900~keV). The three stacked histograms refer to three different ranges for $f$.}
    \label{fig:PLindex_V_vs_energy}
\end{figure}

By assuming, for each energy channel, the geometric mean $E$ of its boundary values and fitting the three points of each GRB and each combination of $f$ and $\beta$ values with a power-law, $V\propto E^{\alpha}$, we obtained an acceptable fit for 82\% of cases. The corresponding $\alpha$ distribution is shown in Figure~\ref{fig:PLindex_V_vs_energy} for three different ranges of $f$. Most values differ from zero by a relatively small amount, with a median value $\alpha_{\rm med} = 0.13$ and $[0.0, 0.24]$ as interquartile range, when all values of $f$ are included.

We used this result to study the impact of using the two measures of $V_f$ obtained with the full passbands of both BAT and GBM interchangeably:  we derived the distribution of the ratio $\xi$ between $V_f$ calculated at 47~keV and $V_f$ calculated at 85~keV, corresponding to the geometric mean energy of the 15--150 and of the 8-900~keV passbands, respectively. The median value is $\xi_{\rm med}=1.08$, with $0.82$ and $1.24$ as the 5\% and 95\% quantiles, respectively. This can be summarised as a $\lesssim 20$\% discrepancy for most measures of $V_f$ obtained with BAT and with GBM.

\section{Comparison and test with past results}
\label{app:compat_G05}
As anticipated in Section~\ref{sec:comp_past}, we concluded that the correlation between $L_{\rm iso}$ and $V_f$ that was found in the early years by \citetalias{Reichart01} and confirmed by \citetalias{Guidorzi05b} was possibly an artefact caused by the poor sampling of the $L_{\rm iso}$--$V_f$ space. 

\begin{figure}
   \includegraphics[width=0.47\textwidth]{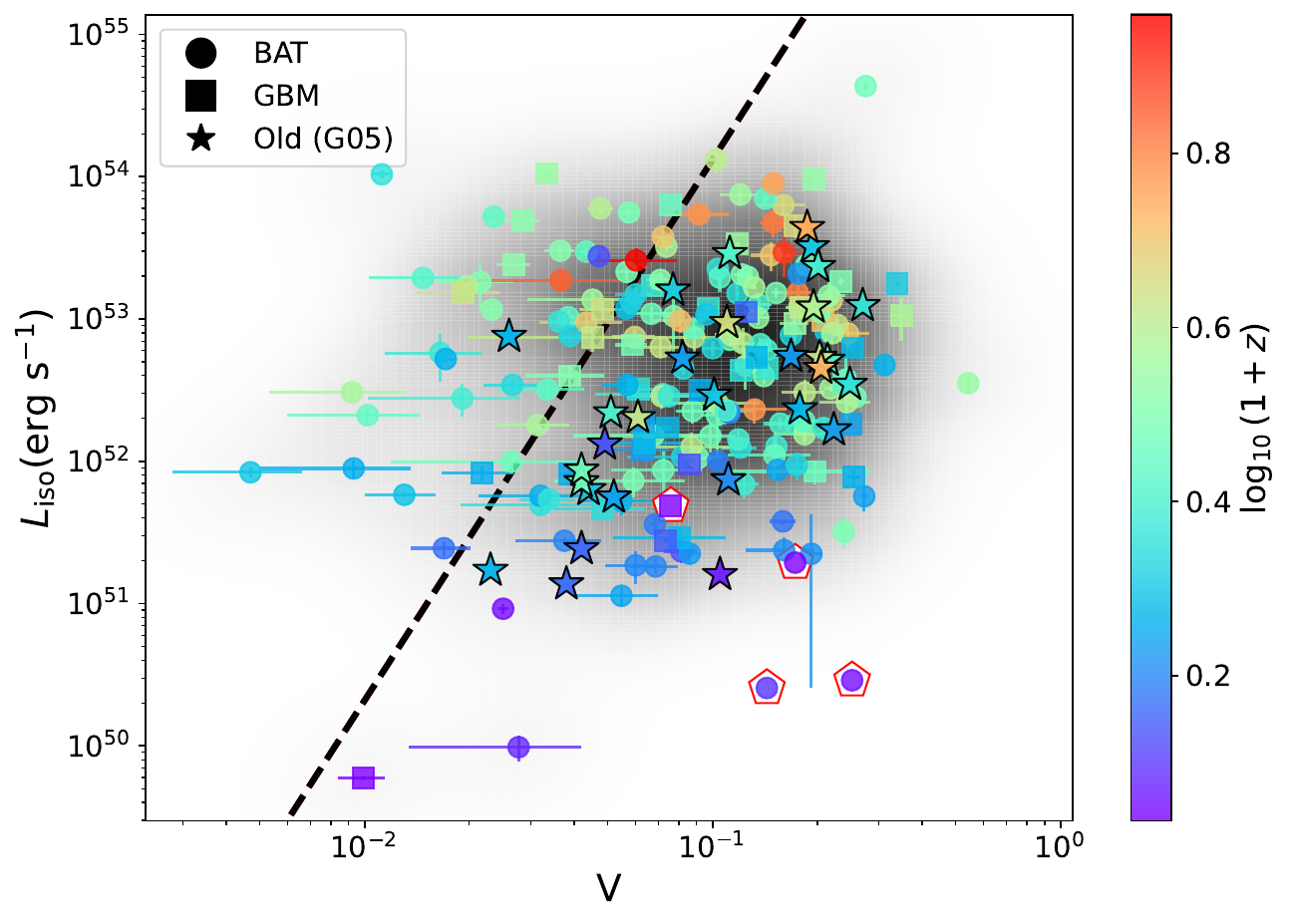}
    \caption{Same as Figure~\ref{fig:VL_f0.45_beta0.6_z} with an additional dashed line that shows the region below which all but one old points lie.}
    \label{fig:V_L_sep_line}
\end{figure}

Here we investigate the reason why the evidence for correlation was stronger for the smaller sample of \citetalias{Guidorzi05b}: this is easily understood by looking at Figure~\ref{fig:V_L_sep_line}: in the region with low $V_f$ and high $L_{\rm iso}$ there is only one old GRB (000210). It is possible to identify a power-law (dashed line) below which all old GRBs but 000210 lie: $L_{\rm iso} = (3\times 10^{49}\ {\rm erg\ s}^{-1})\,(V_f/0.006)^{3.8}$. 
The slope of this power-law, $3.8$, is also consistent with the slope $3.3_{-0.9}^{+1.1}$ that was found by \citetalias{Reichart01} to describe the relation between $L_{\rm iso}$ and $V_f$.
Assuming that the distribution of the old \citetalias{Guidorzi05b} GRBs in the $V_f$--$L_{\rm iso}$ space is the same as the one of BAT and GBM GRBs, we estimated the 
probability that all of the 25, except for one at most, lie on the same side of the power-law by accident: 42/184 from the joint BAT-GBM sample lie above the boundary power-law. Assuming $p=42/184=22.8$\% as the probability for a single GRB to lie on the left side, the corresponding odds that at least 24 out of 25 lie on the same side of the dividing line are easily calculated with a two-sided binomial test\footnote{We used the {\tt scipy.stats.binomtest} function.} and are 3\%, so not impossible. Also, considering that the dividing line was found a posteriori, the correct odds are likely somewhat greater. In fact, a Kolmogorov-Smirnov 2-D test\footnote{The function {\tt ks2d2s} from \citet{NumRecC} was used.} in the $\log{V_f}$--$\log{L_{\rm iso}}$ plane between our set and the \citetalias{Guidorzi05b} one yields a p-value of 21\%, which confirms the absence of evidence for a different parent distribution. 

We conclude that stronger evidence for the $V_f$--$L_{\rm iso}$ correlation that was found in the early years was the result of a poor sampling of the $V_f$--$L_{\rm iso}$ plane.

\end{appendix}

\end{document}